\documentclass[preprint2]{aastex}

\usepackage{graphicx}
\usepackage{epsfig}
\usepackage{times}
\usepackage{natbib}
\usepackage{url}
\usepackage{color}
\usepackage{amssymb,amsmath}
\usepackage{hyperref}
\shorttitle{-}
\shortauthors{ }

\DeclareUnicodeCharacter{2212}{-}

\begin{document}

\title{Magnetic Reconnection between a Solar Jet and a Filament Channel}

\author{Garima Karki\altaffilmark{1}, Brigitte Schmieder \altaffilmark{2,3,4}, Pooja Devi \altaffilmark{1}, Ramesh Chandra \altaffilmark{1}, Nicolas Labrosse \altaffilmark{4}, Reetika Joshi \altaffilmark{5,6}, Bernard Gelly \altaffilmark{7}}
 
\affil{$^1$ Department of Physics, DSB Campus, Kumaun University, Nainital 263 001, India \email{garimakarki31@gmail.com}}
\affil{$^2$ Observatoire de Paris, LIRA, UMR8109 (CNRS), F-92195 Meudon Principal Cedex, France}
\affil{$^3$ Centre for mathematical Plasma Astrophysics, Dept. of Mathematics, KU Leuven, 3001 Leuven, Belgium} 

\affil{$^4$  SUPA, School of Physics \& Astronomy, University of Glasgow, Glasgow G12 8QQ, UK}

\affil{$^5$ Rosseland Centre for Solar Physics, University of Oslo, P.O. Box 1029 Blindern, N-0315 Oslo, Norway}
\affil{$^6$ Institute of Theoretical Astrophysics, University of Oslo, P.O. Box 1029 Blindern, N-0315 Oslo, Norway}
\affil{$^7$ THEMIS, Tenerife,ES}

\begin{abstract}
The solar corona is highly structured by bunches of magnetic field lines forming either  loops, or twisted  flux ropes representing prominences/filaments, or very dynamic structures such as jets. The  aim of this paper is to understand the interaction between filament channels and jets. We use high-resolution H$\alpha$ spectra obtained by the ground-based telescope T\'elescope H\'eliographique pour l’Etude du Magn\'etisme et des Instabilit\'es Solaires (THEMIS) in Canary Islands, and data from Helioseismic Magnetic Imager (HMI) and Atmospheric Imaging Assembly (AIA)  aboard the Solar Dynamics Observatory (SDO). In this paper we present  a multi-wavelength study of the interaction of  filaments and jets. They both consist of cool plasma embedded in magnetic structures. A  jet is particularly well studied in all the AIA channels with a flow reaching 100--180 km s$^{-1}$. Its origin is linked to cancelling flux at the edge of the active region. Large Dopplershifts  in H$\alpha$ are derived  in a typical area for a short time (order of min). They correspond to flows around 140 km s$^{-1}$. In conclusion we conjecture  that these flows correspond to some interchange of magnetic field lines between the filament channel and the jets leading to cool plasmoid ejections or reconnection jets perpendicularly to   the jet trajectory.
\end{abstract}



\section{Introduction}
     \label{S-Introduction} 
The interaction of filaments and jets,  which both  are cool plasma structures in the corona is rarely observed. Many studies concern individually jets and filaments.

Solar jets are  ubiquitous phenomena on the Sun, defined as the beam-like plasma ejections along  open or slightly oblique magnetic field lines, from the lower-to-upper solar atmosphere. Jets are source of sufficient mass and energy input to the higher solar atmosphere, and can be responsible for the heating of the solar corona and acceleration of the solar wind \citep{Raouafi2016, Shen2021, Schmieder2022}. Jets can occur in various regions of the Sun, like active regions (ARs), coronal holes (CHs) and quiet Sun regions, and are often associated with micro-flares and coronal bright points \citep{Schmieder2013, Shen2021}. Jets are frequently observed across a broad range of wavelengths from optical range, called frequently surges  \citep{Roy1973,Schmieder1983}, and  EUV  jets \citep{Schmieder1984} to X-ray jets observed with the solar maximum mission (SMM) XRT instrument  \citep{Schmieder1988, Schmieder_Sim1996} and  with  the Yohkoh mission \citep{Shibata1992, Schmieder_Shi1996, Canfield1996}  throughout the entire solar cycle. Since these pioneer missions jets have frequently been observed in H$\alpha$ and H$\beta$ \citep{Liu2022,Cai2024,Joshi2024} using high  spatial resolution ground-based instruments  like Goode Solar Telescope (GST),  Swedish Solar Telescope (SST), in UV with Transition Region and Coronal Explorer (TRACE)  \citep{Alexander1999} and since 2010 with the Atmospheric Imaging Assembly (AIA: \citealt{Lemen2012}) aboard the Solar Dynamics Observatory (SDO) \citep{Sterling2015, Chandra2015,Joshi2017, Joshi_Gui2020, Shen2021, Schmieder2022}. Their measured height, width, speed, and lifetime range from 1--50 $\times$ 10$^4$ km, 1--10 $\times$ 10$^4$ km, 100--500 km s$^{-1}$, and seconds to sometimes an hour, respectively \citep{Shimojo1996, Nistico2009, Paraschiv2015, Raouafi2016, Shen2021}.

Surges, often observed in H$\alpha$, are considered the cooler counterparts of solar jets and frequently accompanied with hot jets \citep{Roy1973, Schmieder1983, Canfield1996, Liu2004, Jiang2007, Uddin2012, Chandra2015, Nobrega-Siverio2021}. 
Depending on the wavelengths in which they are observed jets have been classified as X-ray jets, extreme ultraviolet (EUV) jets, and H$\alpha$ jets. Jets are classified as straight anemone jets and two sided loop jets according to their morphology \citep{Shibata1994}.
Based on the Hinode X-ray observations solar jets can be classified broadly into two types namely: standard and blowout jets \citep{Moore2010, Sterling2022}. Usually blowout jets are  eruptive jets and can be associated with  mini-filament eruptions and sometimes with coronal mass ejections (CMEs) \citep{Sterling2015, Chandra2017,Joshi_CME2020}. On the other hand the standard jets are not associated with any kind of eruption.  Standard jets are characterized by a narrow spire and a relatively dim base. In contrast, blowout jets initially resemble standard jets with a narrow spire but differ by having a large bright base. Subsequently, blowout jets undergo  violent flux rope eruption, leading to a significant broadening of the spire.


\begin{figure*}[!t]
    \centering
    \includegraphics[width=\textwidth]{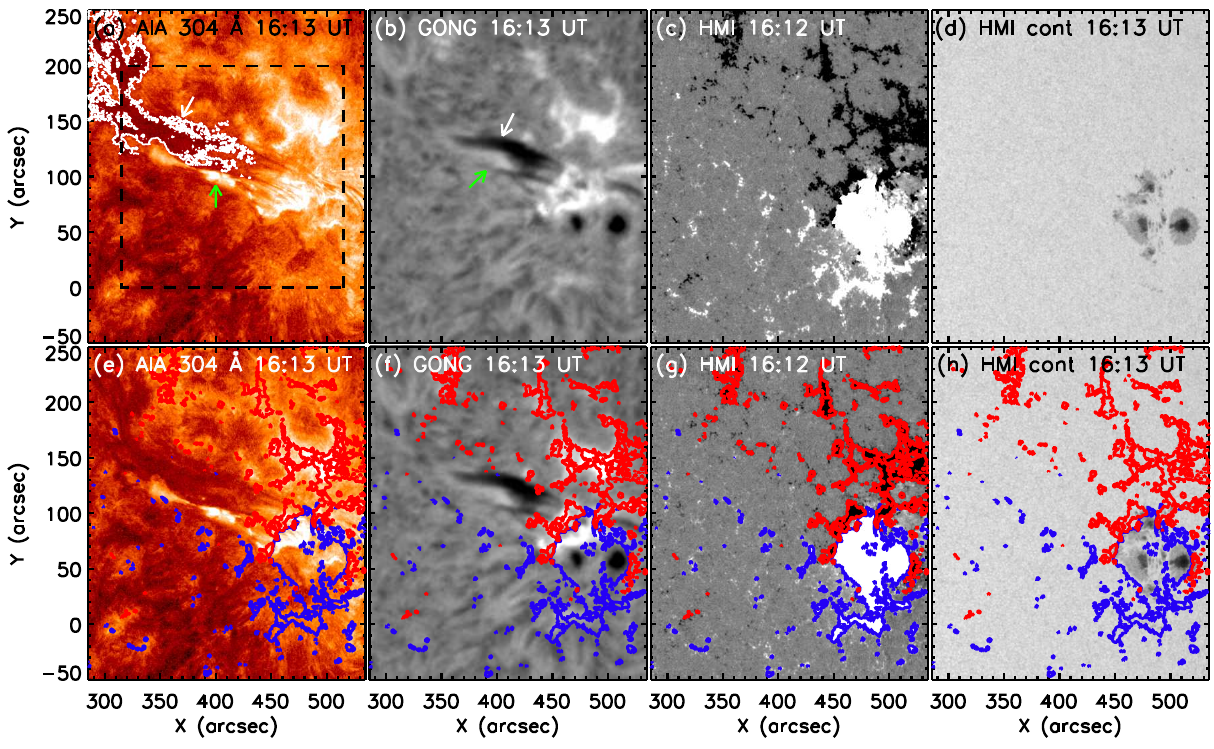}
    \caption{Solar jet and filament observed in AIA 304~\AA\ (panels a, e) and GONG H$\alpha$  (panels b,  f) on 2023 September 25. HMI magnetograms of NOAA AR 13435 (panels c, g). The red and blue contours in panels e-h are for negative and positive polarities, respectively, with a level of $\pm$ 100 G. The sunspots observed in HMI continuum images are shown in panels d, h. The white and green arrows show the positions of the filament and jet, respectively. The dashed box in panel a shows the field of view of Figure~\ref{fig:jet_evolution_difference}. Panels a-d only differ from panels e-h by the presence of contours showing polarities in the latter.}
    \label{fig:aia_gong_hmi}
\end{figure*}

 \begin{figure*}[!t]
     \centering
     \includegraphics[width=0.8\textwidth]{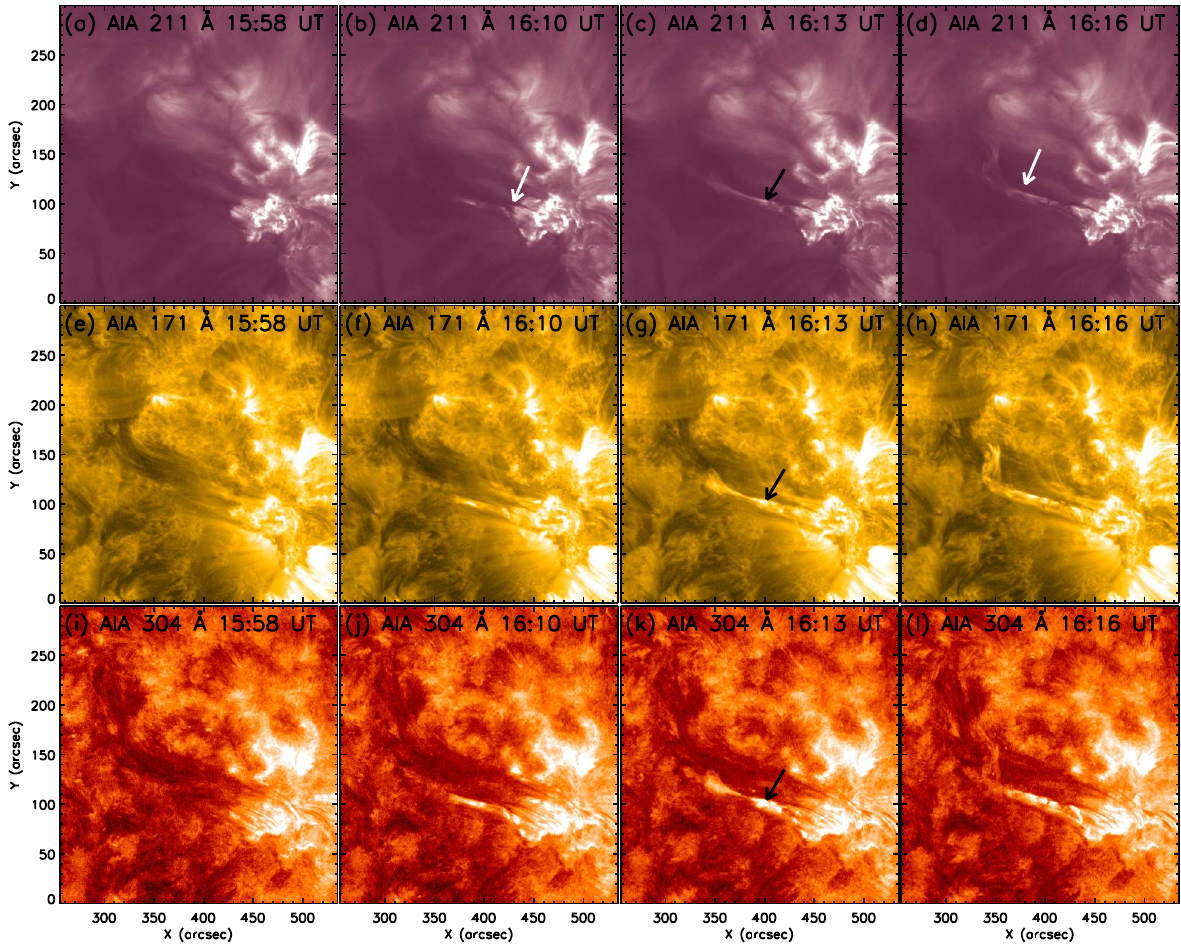}
     \caption{Evolution of the filament observed in absorption  in AIA 211, 171 and 304~\AA. Long threads are resolved. They originate in the bright region and are very dynamic. The bright jet is observed parallel to the filament threads. The white arrow in panels b and d points the dark thread in AIA 211~\AA\ indicating  absorption by filament threads. The black arrow in panels c, g, k shows the brightening corresponding to the red and blue ovals in Figure~\ref{fig:GONG_THEMIS}. An animation of this figure is available. The animation starts at 15:58 UT and ends at 16:16 UT. The real time duration of the animation is 15~s.}
     \label{fig:211evolution}
 \end{figure*}

\begin{figure*}[!t]
\centering
	\includegraphics[width=0.8\textwidth]{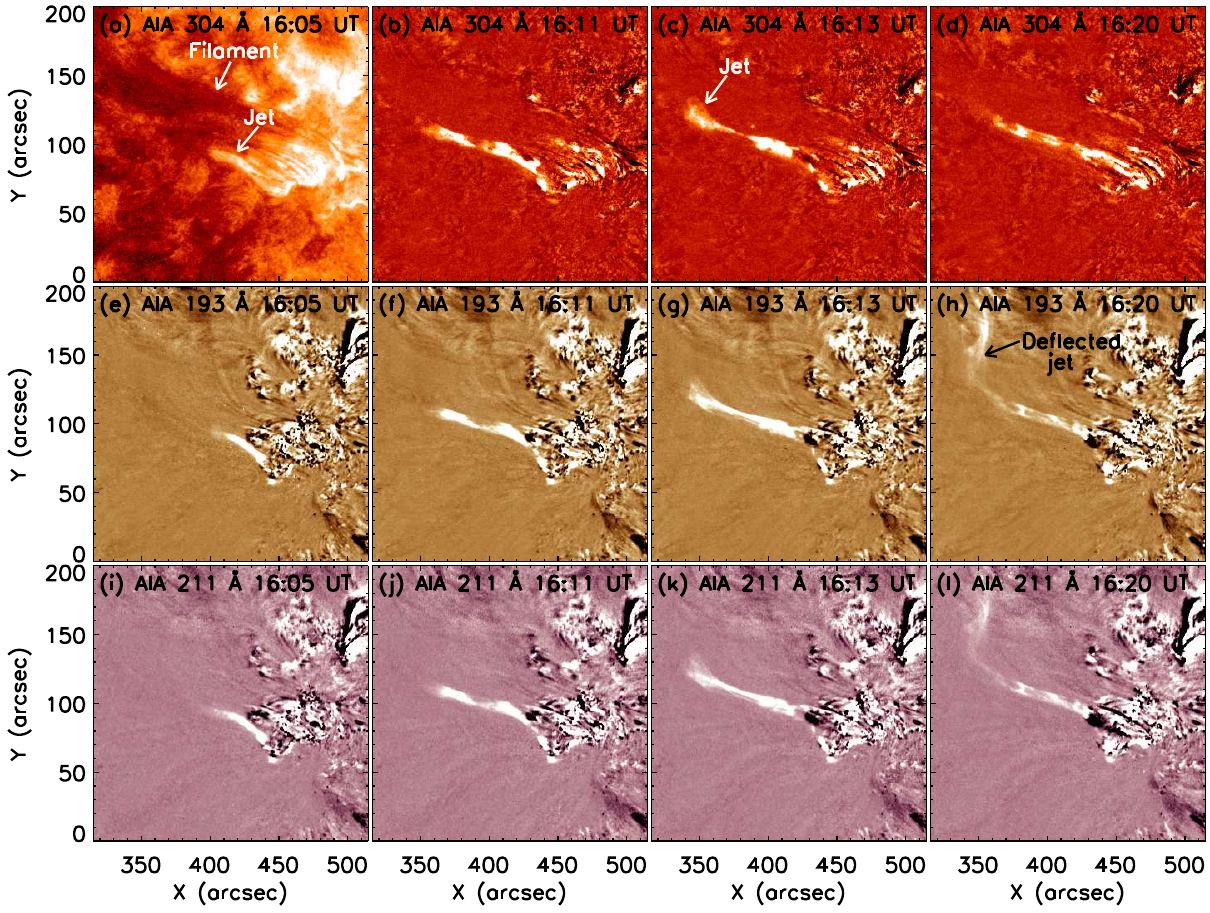}
    \caption {Evolution of the jet  in different AIA wavebands. The panel a shows the intensity image of AIA 304~\AA, panels b--l show base difference images (base time: 15:45 UT) of AIA filters at 304~\AA, 193~\AA \ and 211~\AA, respectively. The arrows in panel c and h indicate the jet leading edge and deflected jet, respectively. An animation is attached with this figure, which starts at 16:01 UT and ends at 16:34 UT. The real time duration of the animation is 28~s.}
    \label{fig:jet_evolution_difference}
\end{figure*}


Magnetic reconnection is believed to trigger the onset of solar jets. The magnetic reconnection process involves the reconfiguration of the magnetic field lines. Three primary scenarios have been investigated for the reconnection of magnetic field lines, which drives solar jets:  flux emergence, flux cancellation, and the loss of equilibrium or instability onset \citep{Schmieder2022}. 

In the flux emergence scenario, reconnection occurs between the newly emerged magnetic flux from beneath the photosphere and the ambient open magnetic field lines and triggers the jet \citep{Nobrega-Siverio2017,Nobrega-Siverio2021}. The breakout model can trigger the jets \citep{Wyper2018, Wyper2019}. In the magnetic flux cancellation scenario, jets can be triggered by the eruption of small-scale filaments known as mini-filaments. These mini-filaments erupt due to magnetic flux cancellation occurring near the magnetic neutral line and beneath the mini-filament \citep{Sterling2016,Panesar2016a, Moore2022}.  Destabilization  and eruption of mini-filaments can  trigger  confined plasma ejections \citep{Poisson2020}. In the instability onset or loss of equilibrium scenario, jets are initiated when 
non-potential, stressed closed magnetic flux beneath the null point undergoes reconnection with quasi-potential ambient flux exterior to the fan surface \citep{Joshi2024F}. This reconnection process occurs explosively once a critical threshold is reached, often in response to quasi-static footpoint motions \citep{Torok2009,Joshi_Gui2020}. Importantly, in this scenario, the reconnection is spontaneously initiated by internal rearrangements within the coronal field of the closed flux system. There is typically no change in the amount of unsigned vertical magnetic flux in or near the jet source during this process.

Solar filaments, also called solar prominences when viewed at the limb, are cool plasma structures embedded in the hot, million degree corona. 
They are also much denser  by two orders of magnitude than their environment and are supported by magnetic field against gravity \citep{Labrosse2010, Mackay2010}. Filaments are observed directly as dark features in chromospheric lines by their contrast with the chromosphere emission. They are also visible  as dark features in the AIA filters at  171~\AA, 193~\AA\ and 211~\AA\ due to absorption  of the EUV emission by the cool plasma  \citep{Heinzel2001,Schmieder_Lin2004, Labrosse2011}. Filament channels lie between positive and negative polarities. The filament threads make a small angle ($<$ 25$^{\circ}$) with the polarity inversion line (PIL). The H$\alpha$  filament is supported in the dips of magnetic field lines \citep{1998Aulanier1998}. The magnetic field lines are forming flux rope or sheared arcades \citep{Aulanier_De2002,Guo2010}. Both contribute to represent  the filament channel which is well visible in EUV \citep{Aulanier_BS2002}. The EUV filament,  well visible in AIA 304~\AA\, can be broader than the  H$\alpha$ filament by a factor 5~\citep{Heinzel2001}.

Magnetic reconnection is a frequent process due to the bunches of magnetic field lines in the corona. It is also one way to explain  coronal heating, when multiple brightenings are  observed with Solar Orbiter  \citep{Berghmans2021, Panesar2021, Zhukov2021}. Recently magnetic reconnection has been evidenced by spectroscopy observations. Using IRIS Mg II spectra, bilateral flows at the reconnection point have been evidenced. This could initiate a jet in the corona \citep{Ruan2019,Joshi_Gui2020} or heating of jet threads \citep{Cai2024} or developed  mini-jets in a braiding loop \citep{Antolin2021}. Such mini-jets or  jets of reconnection show  flows in both directions. They concern a few pixels and last a short time ($<$ min). In the corona many magnetic structures  are present and can interact  such as loops, jets, filaments.
 Recent studies show that the interaction of jets and filaments can lead to heating or oscillations in the filament \citep{Luna2021, Joshi2023, Cai2024,Luna2024}.

In this paper, a multi-wavelength study of   recurrent jets observed with  SDO/AIA, HMI, and ground-based THEMIS instruments  reveal  their  origin and their characteristics  (Section 2). We discuss the presence of a filament with a large EUV filament channel in the vicinity of the jets (Section 3). The paper  focuses on one of the jets observed by THEMIS. The aim of the paper is to localize the areas  in the jet where high   Dopplershift velocities of  cool plasma  are detected    in the  H$\alpha$ spectra of THEMIS\  (Section 4). We interpret these fast  flows as a  spectroscopic signature of   the interchange of magnetic field lines between the jet and the filament leading to  magnetic reconnection (Section 5).

\section{Instruments, Data Sets, and Methods}  
\subsection{Instruments} 

For this study we have used the data from the following instruments.

The Atmospheric Imaging Assembly (AIA: \citealt{Lemen2012}) aboard the Solar Dynamics Observatory (SDO, \citealt{Pesnell2012}) provides full disk images of the Sun's surface in several wavebands, offering insights into different layers of the solar atmosphere. These wavebands include seven EUV wavelengths: 94~\AA\ (6 MK), 131~\AA\ (10 MK), 171~\AA\ (600,000 K), 193~\AA\ (1 MK), 211~\AA\ (2 MK), 304~\AA\ (50,000 K), and 335~\AA\ (2.5 MK), two UV wavelengths, 1600~\AA\ (10,000 K) and 1700~\AA\ (4500 K), and one white-light wavelength: 4500~\AA\ (6000 K).  AIA observes with a cadence of 12 s for EUV, 24 s for UV, and 3600 s for white-light, all with a pixel size of  0.6\arcsec. In this work, we used AIA data specifically from the 171~\AA, 193~\AA, 211~\AA, and 304~\AA\ wavebands to analyze solar jets across different layers of the Sun's atmosphere.
The Helioseismic Magnetic Imager (HMI: \citealt{Schou2012}) aboard SDO observes the entire solar disk with a pixel size and temporal resolutions of 0.5\arcsec and 45 s respectively, at 6173~\AA. For the current study, we used line-of-sight (LOS) magnetic field maps, and white-light data from HMI to analyze the magnetic structure and distribution of sunspots within the AR.

The T\'elescope H\'eliographique pour l’Etude du Magn\'etisme et des Instabilit\'es Solaires (THEMIS: \citealt{Mein1985}), features an helium filled Cassegrain telescope equipped with a Ritchey-Chretien primary mirror of 92 cm diameter, installed on top of  a 22.5 m height observing tower. The front and back parts of the telescope are closed by glass window plates, hence sealing the telescope to avoid turbulence above the main mirror. A 9-meter diameter dome shields the instrument from environmental elements. The telescope is equipped with highly flexible spectroscopic capabilities that enable simultaneous observation of several spectral ranges that can be strategically distributed over a 2500~\AA\ bandwidth at observer’s choice. This unique capability is made possible using first a grating predisperser featuring three interchangeable gratings, and then an echelle spectrograph. Spectra are acquired using either recent CMOS cameras (2k $\times$ 2k) and/or three older EMCCD cameras (512 $\times$ 512), with a maximum of five spectral ranges allowed. The telescope offers simultaneous high quality imaging along with the multi-line spectroscopy (MTR2) mode. For the current study we have used the THEMIS H$\alpha$ spectra obtained on the larger cameras. The pixel resolution along the slit is 0.06\arcsec. The slit width is 0.5\arcsec and the stepping is either 0.5\arcsec or 1$\arcsec$. The spectral dispersion is $\sim$ 3.076 m\AA\ per pixel. The bandpass is 6.3~\AA\ around 6563~\AA. Exposure times for H$\alpha$ range from 0.05 to 0.2 s depending on targets and seeing conditions. The adaptive optics (AO) \citep{Tallon2022} was used and allows us to have high-resolution data (except for prominences targets). However, due to the relatively moderate seeing we could not get long sequences of observations. We only obtained a few frames from THEMIS with a step of 1$\arcsec$ and an exposure time of 0.2~s allowing comparison to space-borne observations and GONG.

The Global Oscillation Network Group (GONG: \citealt{Harvey1996}) operates a network of six ground-based observatories strategically located around the world to provide continuous, high-resolution observations of the Sun. These observatories use H$\alpha$ filtergrams to capture full-disk images of the Sun with a temporal and spatial resolution of 1 min and 2$\arcsec$, respectively. Here we have used GONG H$\alpha$ data to observe the surges and filaments on the solar disk. 

\subsection{Methods of Analysis} 
We mainly used AIA 171, 193, 211, 304~\AA\ images to observe the jet structures at chromospheric, transition-region and coronal temperatures. All  images from AIA and HMI are aligned at 15:45 UT using drot\_map in solar software to correct for the solar differential rotation. For the coalignment of GONG H$\alpha$ data with SDO/AIA data, we have compared the centroid of sunspots visible in the GONG H$\alpha$ and SDO/HMI continuum images at the same time (Figure \ref{fig:aia_gong_hmi} panels b and d). To study the jet evolution movies with a cadence of 12 s are created. 

For the spectroscopic analysis we analysed the THEMIS H$\alpha$ spectra using SSWIDL routines. We plotted the spectra for each slit position and find the slit positions corresponding to maximum plasma flows. For these slit positions we plotted the spectral profiles and found the Dopplershifts using the relation: velocity = $\frac{\Delta \lambda}{\lambda}c $.

\begin{figure}[!t]
    \centering
\includegraphics[width=0.49\textwidth]{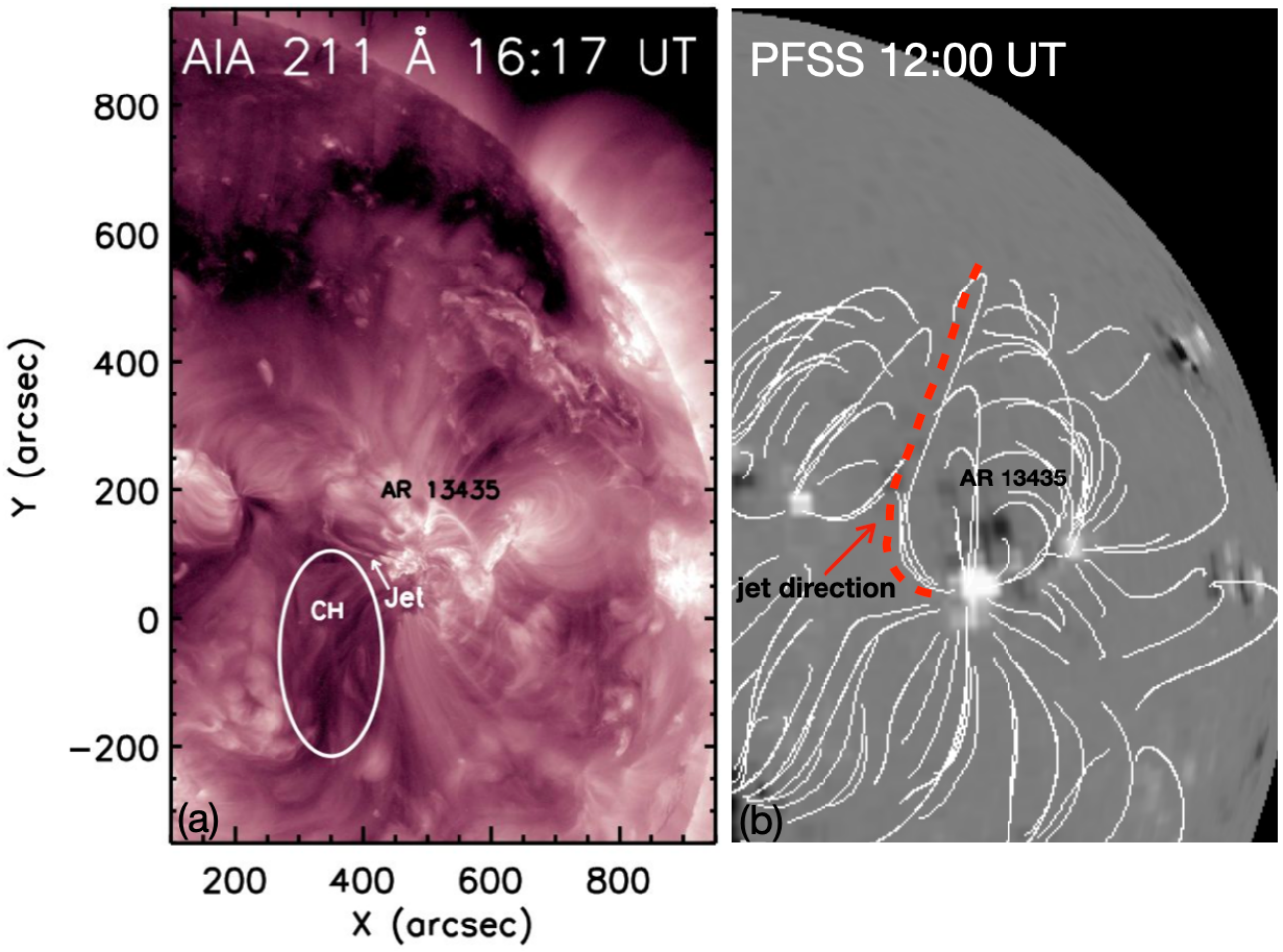}
    \caption{Left panel:  Small CH in the vicinity of the jet observed in AIA 211~\AA. Right panel: PFSS extrapolation in the jet region. Long magnetic field lines originating from the East of AR 13435 are going to the North, showing the path of the jet flow. The possible jet flow path is shown with the red dashed line.}
    \label{fig:pfss}
\end{figure}

\section{Jet and Filament Characteristics}
\subsection{Overview} 
 The NOAA AR 13435 located at N10W37 is a jet productive AR on 2023 September 25. Between 00:00 UT to 23:59 UT, the AR produced around 24 multiple jets in the same region, separated by 30 minutes to one hour  and with variable lengths (20--80 Mm) (Table \ref{tab:Jets}). The jet that is observed by THEMIS is the 16$^{th}$ jet (Jet 16) at 16:01 UT. The previous jet (Jet 15)  occurs  at 15:17 UT. Jet 16  is accompanied by  a jet  clearly  visible in all AIA wavelengths almost simultaneously around 16:01 UT.
 \begin{table*}[!t]
     \centering
      \caption{Jets observed on 2023 September 25 from NOAA AR 13435.}
      \vspace{0.3cm}
     \begin{tabular}{cccc|cccc}
     \hline
      Jets    &  Onset time & Location & Max length  & Jets & Onset time & Location & Max length \\
          &  (UT) & (arcsec) & (Mm) &  & (UT) & (arcsec) & (Mm) \\
      \hline
        Jet 1 & 00:35  & (450,49) & 53 & Jet 13 & 13:45 & (449,61) & 22 \\ 
       Jet 2 & 00:56 & (448,45) & 39 & Jet 14 & 14:00 & (454,68) & 57 \\ 
       Jet 3 & 01:28 & (447,48) & 51 & Jet 15 & 15:17 & (453,
       66) & 25 \\
       Jet 4 & 02:01 & (477,71) & 22 & Jet 16 & 16:01 & (450,62) & 73 \\
       Jet 5 & 02:47 & (447,44) & 60 & Jet 17 & 16:48 & (449,63) & 23 \\
       Jet 6 & 03:02 & (446,39) & 33 & Jet 18 & 17:48 & (450,63) & 20 \\
       Jet 7 & 04:33 & (459,66) & 28 & Jet 19 & 18:12 & (450,63) & 64 \\
       Jet 8 & 09:28 & (454,68) & 29 & Jet 20 & 19:19 & (450,64) & 76 \\
       Jet 9 & 10:32 & (457,67) & 62 & Jet 21 & 19:35 & (450,62) & 23 \\
       Jet 10 & 11:00 & (450,65) & 19 & Jet 22 & 21:42 & (444,61) & 45 \\
       Jet 11 & 12:12 & (448,61) & 21 & Jet 23 & 22:02 & (448,67) & 27 \\  
       Jet 12 & 12:30 & (451,65) & 66 & Jet 24 & 23:32 & (445,62) & 33 \\
          \hline  
     \end{tabular}
     \label{tab:Jets}
 \end{table*}

\begin{figure}[!t]
\centering
\includegraphics[width=0.49\textwidth]{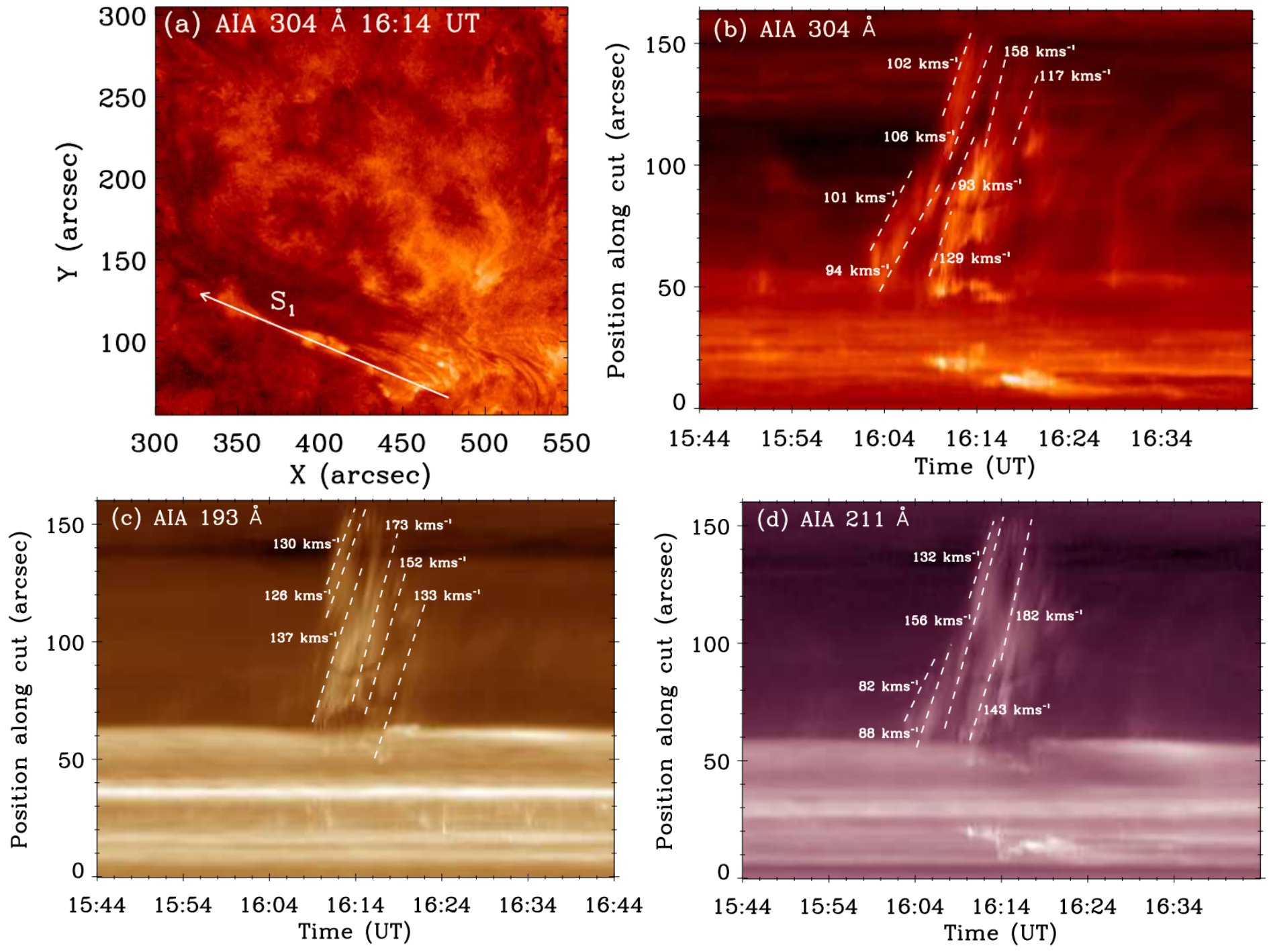}
    \caption {Time slice analysis of the jet at 16:14 UT. Panel a: position of  slit along the jet, panels b-d show the time-distance diagrams along slit S$_1$ and the computed  plane-of-sky speeds at different wavelengths.} 
    \label{fig:time_slice_S1}
\end{figure}

\begin{figure}[!t]
\centering
\includegraphics[width=0.49\textwidth]{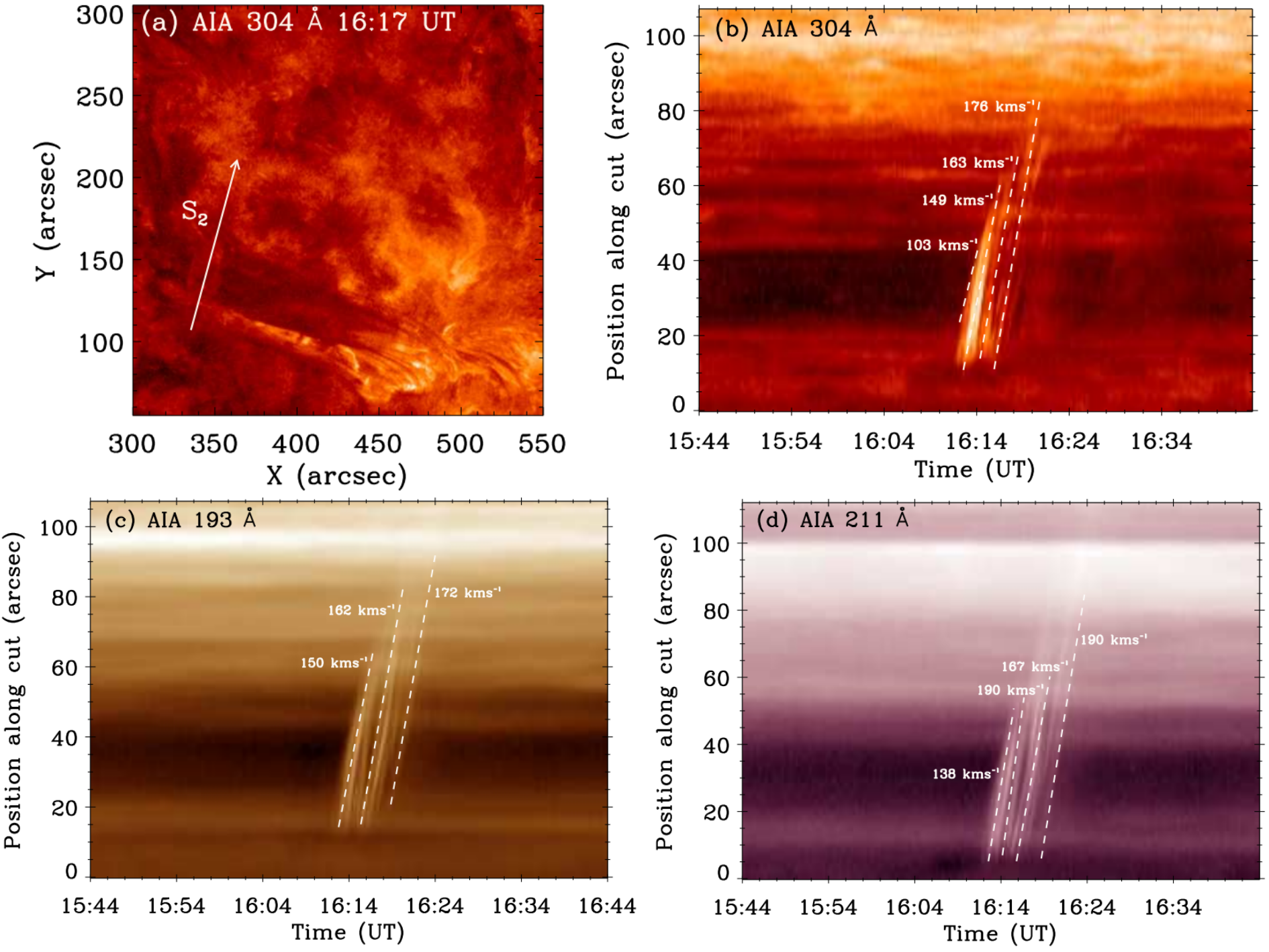}
    \caption {Time slice analysis of the jet at 16:17 UT. Panel a: position of  slit along the jet, panels b-d show the time-distance diagram along slit S$_2$ and the computed  plane-of-sky speeds at different wavelengths.} 
    \label{fig:time_slice_S2}
\end{figure}

\begin{figure}[!t]
\centering
	\includegraphics[width=0.49\textwidth]{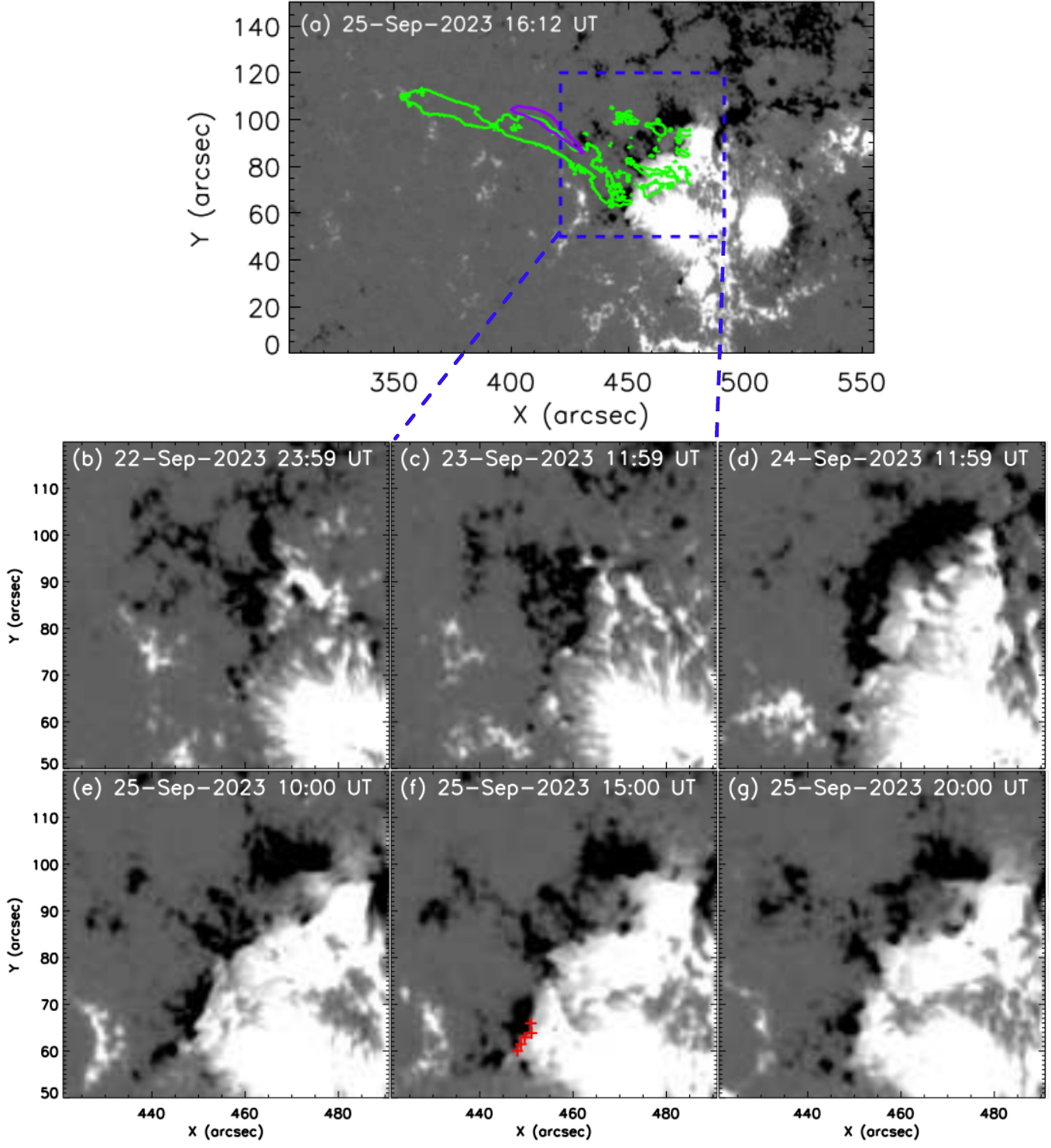}
    \caption{Magnetic field evolution of NOAA AR 13435. Panel a shows an HMI magnetogram image on 2023 September 25 at 16:12 UT. The green, violet contours correspond to the jet observed in AIA 304~\AA, GONG H$\alpha$ respectively. The evolution of the magnetic field with the enlarged view of blue box  drawn in  panel a  is shown in panels b-g. The red crosses in panel f show the location of flux cancellation, corresponding to the jet base.  An animation of this figure is attached. The animation starts on September 22 at 23:59 UT and ends on September 25 at 20:57 UT. The real time duration of the animation is 3 min 50 s.}
    \label{fig:magnetic_field_evolution}
    \end{figure}

\begin{figure}[!t]
    \centering
    \includegraphics[width=0.49\textwidth]{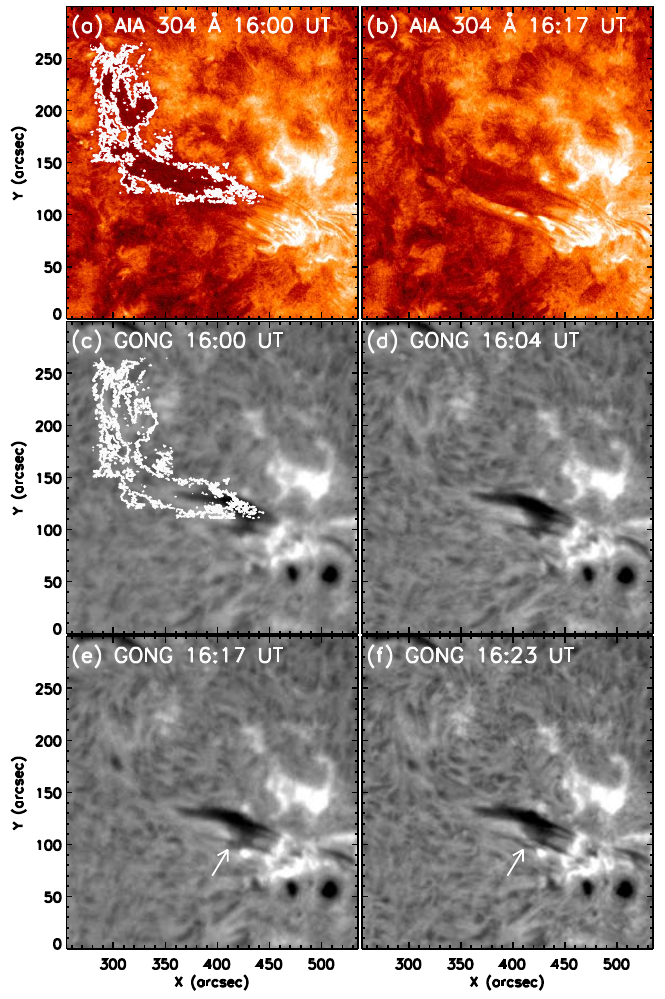}
    \caption{Filament channel and jet observed in AIA 304~\AA\ and GONG H$\alpha$ on 2023 September 25 before the jet (panels a and c) and during its maximum extension (panels b and e, f), respectively. The white contour represents the filament channel  observed in AIA 304~\AA. The white arrow in panels e and f indicates the cool jet (surge). An animation of this figure which starts at 15:59 UT and ends at 16:34 UT is available. The real time duration of the animation is 6~s.} 
    \label{fig:aia_filament_sunspot}
\end{figure}

\begin{figure}[!t]
    \centering
    \includegraphics[width=0.49\textwidth]{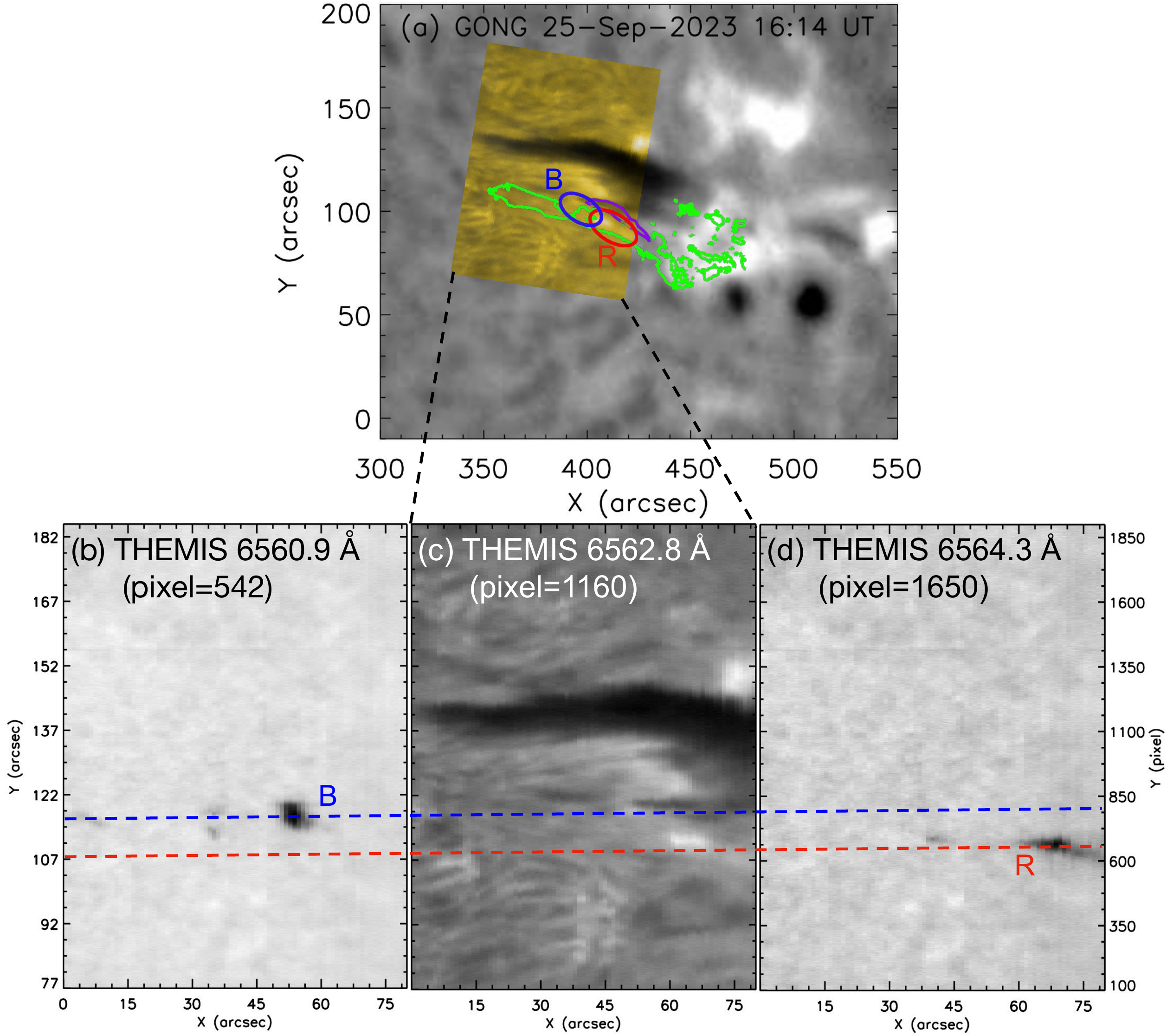}
    \caption{ 
    Panel a displays the GONG H$\alpha$ image at 16:14 UT, overlaid with the THEMIS image derived from the H$\alpha$ spectra highlighted in yellow, and jet contours in AIA 304~\AA\ and GONG H$\alpha$ shown in green and violet, respectively. Panel b exhibits the THEMIS image corresponding to the wavelength 6560.9~\AA\ (blue wing). Panel c presents the H$\alpha$ center image at 6562.8~\AA. Panel d illustrates the THEMIS image at the wavelength 6564.3~\AA\ (red wing).} 
    \label{fig:GONG_THEMIS}
\end{figure}

\begin{figure}[!t]
\centering
	\includegraphics[width=0.49\textwidth]{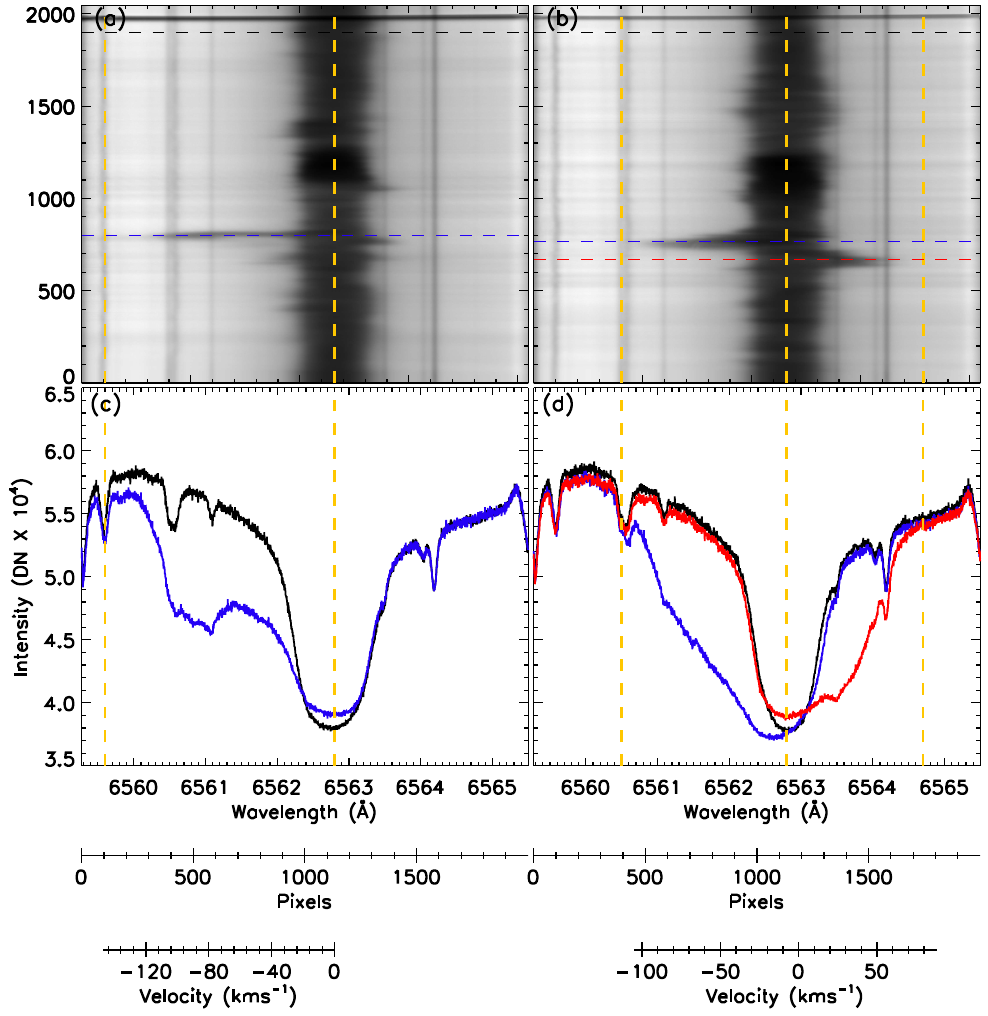}
    \caption{THEMIS H$\alpha$ spectra (top) and profiles (bottom) observed at 16:13 UT for slit 50 (left) and 70 (right). The black 
    line corresponds to the reference profile and the blue and red lines to blue and red shifted profiles, respectively. The vertical yellow lines are at the H$\alpha$ line center and at the maximum shift in the blue and red wings.}
    \label{fig:spectra_16:13}
\end{figure}

\begin{figure}[t!]
\centering
	\includegraphics[width=0.48\textwidth]{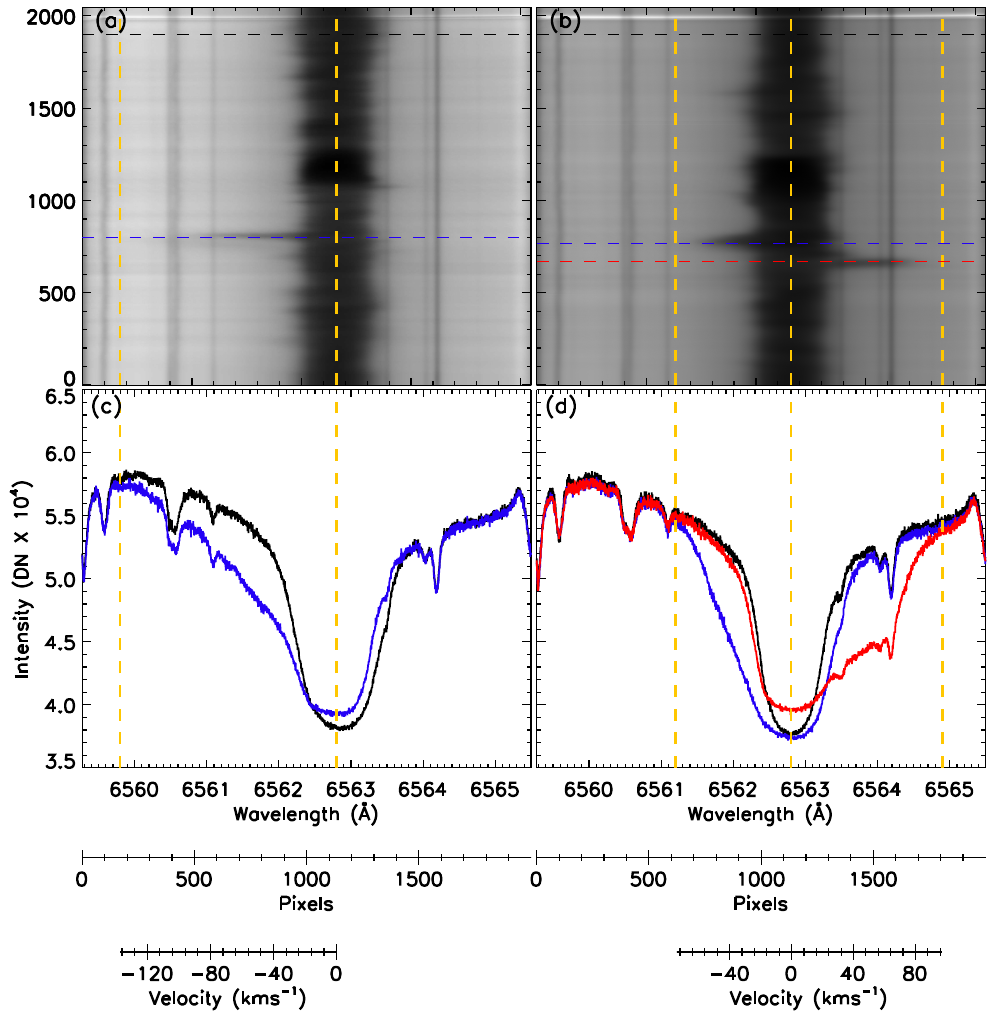}
    \caption{THEMIS H$\alpha$ spectra (top) and profiles (bottom) observed at 16:14 UT for slit 50 (left) and 70 (right). The black 
    line corresponds to the reference profile and the blue and red lines to  blue and red shifted profiles, respectively. The vertical yellow lines are at the H$\alpha$ line center and at the maximum shift in the blue and red wings.}
    \label{fig:spectra_16:14}
\end{figure}

\begin{figure}[!t]
    \centering
    \includegraphics[width=0.49\textwidth]{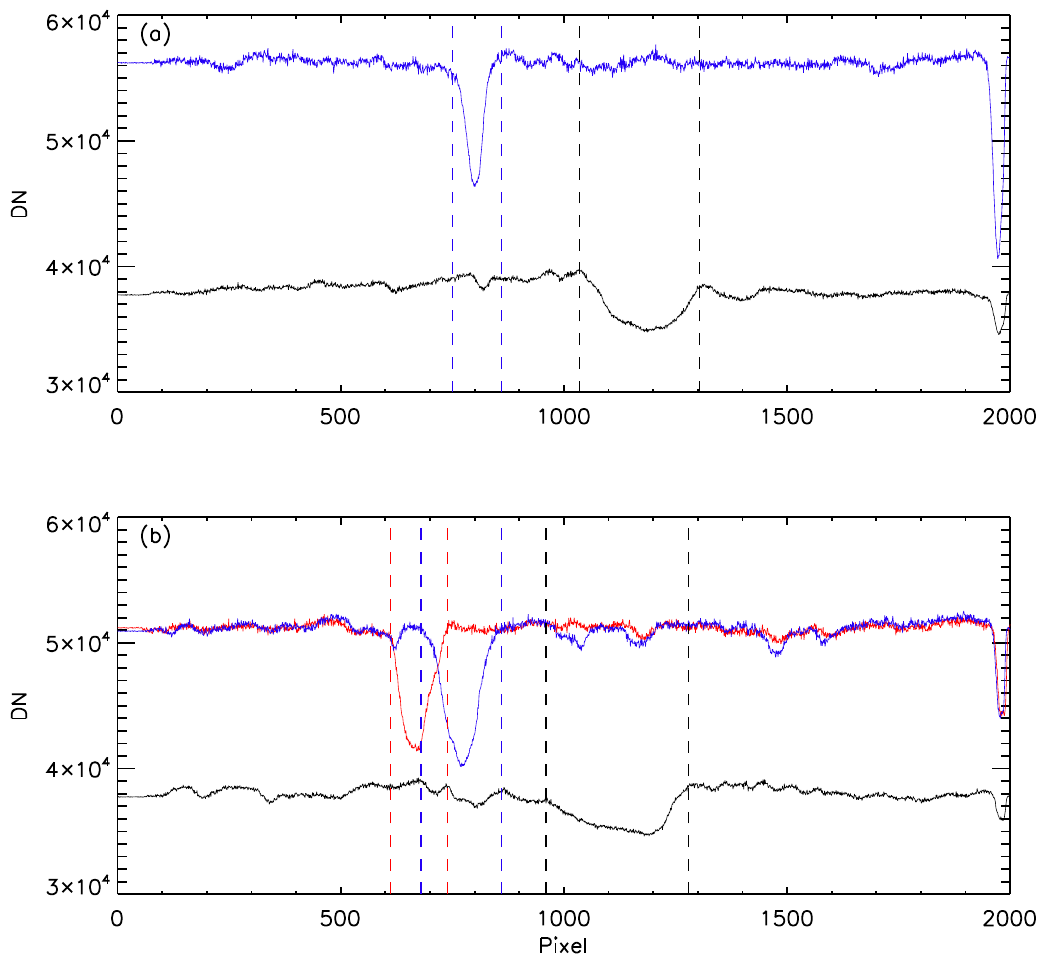}
    \caption{Panel a: Cuts along the slit for position 50  at 16:13 UT in the H$\alpha$ line center (black curve) and in  the  blue wing at  H$\alpha$ -- 2~\AA\ (Pixel 500). Panel b: Cuts along the slit for position 70 in the H$\alpha$ line center (black curve) and in  the  blue wing (blue curve) at  H$\alpha$ -- 0.8~\AA\ (Pixel 900) and in the red wing (red curve) at H$\alpha$ + 0.9 ~\AA\ (Pixel 1450). The dips in the black, blue and red profiles show the locations of  the filament, and of  the fast flows in  the blue and red wings, respectively.}
    \label{fig:curves}
\end{figure}

\begin{figure}[!t]
    \centering
    \includegraphics[width=0.49\textwidth]{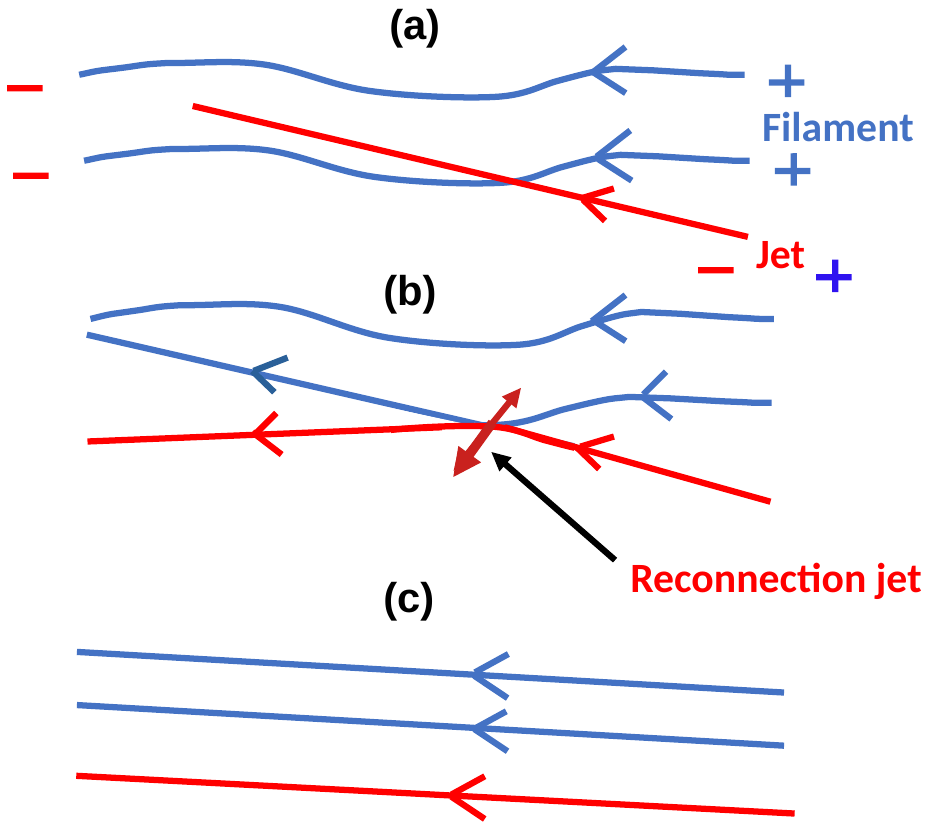}
    \caption{Sketch of the interchange of magnetic field lines between the filament and the jet. Panel a: the blue lines show the filament field lines and the red lines represent the field lines of the jet and its direction. Panel b: reconnection between the blue and red lines creating a reconnection jet (dark red arrows in opposite directions). Panel c: the reconnected magnetic field lines, blue and red, corresponding to the field lines of the filament and  jet respectively.}
    \label{fig:jet_filament_drawing}
\end{figure}

Figure~\ref{fig:aia_gong_hmi} shows the AIA 304~\AA, GONG H$\alpha$ images of the jet called also surge (dark area), HMI magnetogram and HMI continuum image of the AR in the vicinity of a large filament. The surge is at the same location as the 304~\AA\ jet. The large filament observed in H$\alpha$ is also visible in AIA  211, 171, 304~\AA\ as dark elongated features  indicated by white arrow in panels b and d of Figure \ref{fig:211evolution}. This is due to the similarity of the optical thickness of H$\alpha$ and the absorption of the  EUV  continuum  near 193~\AA\ by the cool material \citep{Heinzel2001,Schmieder_Lin2004}.

The evolution of the jet in AIA 304~\AA, 193~\AA, and 211~\AA\  base-difference images from 16:05 UT to 16:20 UT is displayed in Figure~\ref{fig:jet_evolution_difference}.
The filament does not evolve over time so we do not see it in the difference images. Some very fine structures  appear in the north part of the jet which could belong to associated surge, well visible in H$\alpha$ (Figure \ref{fig:aia_gong_hmi}).
The direction of jet ejection is towards the North-East direction, parallel to the filament axis. The maximum projected length  attained by the jet in AIA 304, 193 and 211~\AA\ is almost the same, about 80 Mm. After reaching the maximum length at $\sim$ 16:13 UT, the jet is deflected towards the North direction. This deflection is visible more clearly in AIA 193 and 211~\AA\ base difference images. The angle of deflection is nearly equal to 90$^{o}$. After the deflection, the jet can be visible up to 16:20 UT and the length traveled by jet after the deflection is $\sim$ 58 Mm. We noticed that the CH structure is visible in the East-South of the AR. The location of CH along with the jet is depicted in Figure~\ref{fig:pfss}. For a  better view of the jet evolution, we refer to the accompanying movie. 
To explain the deflection of the jet, we explore the magnetic connectivity of the AR using PFSS extrapolation technique. The result of this extrapolation is shown in Figure~\ref{fig:pfss}. From the extrapolation, we find the turning of the magnetic field lines from the location of the jet deflection, which provides a clear path for the jet ejection. Upon examination of all the jets, we find that  jets attaining heights greater than 33~Mm show  clear deflection, while other jets do not show such deflection. This may be due to the lower height, as these jets could not  be deflected by the nearby CH.

\subsection{Dynamics of the Jet}

We use the time-distance technique to analyse the jet observed in all the AIA channels. 
The plane-of-sky speeds of the jet in various AIA wavelengths are determined by the slope of the brightening in the time-distance diagram for both the initial and the deflected directions of the jet. In the time-distance method, an artificial slit is fitted along the jet ejection direction. Along the slit S$_1$, the jet ejection shows multiple strands/threads with different speeds ranging from 93--158, 126--173 and 82--182 km s$^{-1}$ in AIA 304, 193, 211~\AA, respectively.  Higher speeds are found at the top of the jet. The time-distance plots in various AIA wavelengths along with the slit position, are presented in Figure~\ref{fig:time_slice_S1}. 

Further, we have also computed the plane-of-sky speed of the jet after the deflection with the same procedure mentioned above along the slit S$_2$. The jet's speed after the deflection  range from 103--176, 150--182, 138--190 km s$^{-1}$ in AIA 304, 193, 211~\AA, respectively. The time-distance plots along the deflected jet direction in different AIA wavelengths are illustrated in Figure~\ref{fig:time_slice_S2}. Here again the ejection shows multiple ejections. Such a multiple ejection is also reported in previous observations defined as threads \citep{Schmieder2013}. The speeds of the  hot jets have been reported  by several authors (for example: 
\citealt{Schmieder2013,Panesar2016b, Zhang2023}). 
Their computed values range from 100 to 300 km s$^{-1}$, consistent with our results.

\subsection{Magnetic Field Evolution of the AR} 
To understand the magnetic causes of the jet, the line-of-sight photospheric magnetic field observed by HMI is analysed. The photospheric magnetic field of the AR  on 2023 September 25 is shown in Figure~\ref{fig:magnetic_field_evolution}a. Further, the evolution of the magnetic field during September 22--25 around the jet location is displayed in panels b-g. The magnetic polarities in the AR consist of a strong leading positive polarity and a following dispersed negative polarity. On September 22, we observe the positive polarity surrounded by negative polarity on its north side. The contours of the jet  and some brightening  visible in AIA 304 \AA\ are overlaid in panel a of the figure. On September 23--24,  we observe the emergence of positive and negative polarities around the location of the jet origin. Together with the emergence, we also observe that the positive and negative polarities converged and cancelled each other along the PIL between positive and negative polarities during September 24--25. The leading spot is surrounded by a moat region with moving magnetic features (MMF)  which cancel with part of the negative  following polarity (see the HMI movie). The jet occurs at the null region between positive and negative polarities. Magnetic flux cancellation at the edge of the AR-positive polarity triggers the jet. This  is a common mechanism to trigger jets in ARs 
\citep{Mulay2016,Panesar2016b,Poisson2020}.

\subsection{Filament Channel}
GONG H$\alpha$ images are used to study the chromospheric features (filament and surge) associated with the jet. We observed a filament  with a wide structure and many unresolved threads near the jet region in the GONG H$\alpha$ images  before the jet occurrence (Figure \ref{fig:aia_gong_hmi}). By analysing the GONG images we observed a  thin surge accompanying the jet  nearly overlying some filament threads around 16:13 UT. It is difficult to distinguish both. However, in the GONG movie between blurred images there are sharp images showing a  jet crossing a filament thread with a bright point. The estimated speed  is high ($\sim$ 100--150~km s$^{-1}$) but uncertain. In AIA 304 \AA\ we observe a large filament channel covering the H$\alpha$ filament and the small threads (Figure \ref{fig:aia_filament_sunspot}).
 
 Using GONG images we were able to locate the position of the filament  in THEMIS images by overlying THEMIS images over GONG images. Figure~\ref{fig:GONG_THEMIS} panel a  displays the GONG image overlaid with the THEMIS image. THEMIS observed this field of view over  a too short period of time which did not enable us to see the evolution of the jet and compute its speed. The GONG movie has a   spatial resolution that is too low  to allow us to distinguish a surge from the filament threads.  There is no visible difference in the H$\alpha$ GONG images before and during the jet. It is only thanks to the movie that we can detect the presence of a cool jet  (Figure \ref{fig:aia_filament_sunspot}  c - f).

\section{THEMIS Spectroscopy} 

THEMIS images are reconstructed  from the H$\alpha$ spectra obtained with a slit  which scans a small portion of the sun ($80\arcsec \times 108\arcsec$). 
In the H$\alpha$ maps of THEMIS 
the presence of the filament and threads  is indicated as the minimum of counts at the center of H$\alpha$ (Figure \ref{fig:GONG_THEMIS}).  We have 80 spectra obtained in 16 s with a spatial resolution along the slit of 0.06\arcsec. Looking systematically at all the spectra we discovered that a few of them present  extended wings in the blue  and in the red in specific pixels  at two times (16:13 UT and 16:14 UT).

THEMIS maps obtained in the far wings of H$\alpha$   are used to identify  the locations of fast plasma flows. We obtained  images for H$\alpha$ center (6562.8~\AA), for the blue wing  at H$\alpha$ {-- 1.9~\AA\  (6560.9~\AA)} and at H$\alpha$  + 1.5~\AA\ in the red wing (6564.3~\AA). THEMIS maps  for three different wavelengths are presented in Figure~\ref{fig:GONG_THEMIS} panels b, c, d.  In the far wings of H$\alpha$, dark areas appear  in specific pixels indicating regions with fast flows.
We have analysed the H$\alpha$ spectra presenting rapid flows, principally spectra 50 and 70 in the scan of 16:13 UT and in the scan of 16:14 UT. The spectra and profiles of THEMIS H$\alpha$ (6562.8~\AA) are shown in Figure~\ref{fig:spectra_16:13} and \ref{fig:spectra_16:14} observed at 16:13 and 16:14 UT, respectively. 
In these specific spectra the Dopplershifts are very large. 
In the blue area (spectra 50) the blueshift velocity can reach 140 km s$^{-1}$. In the  red area (spectra 70) we observed bilateral flows ($\pm$ 80 km s$^{-1}$). The cuts along the slit for positions 50 and 70 are shown in Figure \ref{fig:curves}.  The section of the filament is large and covers the filament channel for position 70 (around 300 pixels = 18\arcsec\ ). The large flow areas are narrower (a few arcsec). To detect them we need spectroscopic data with high spatial resolution. It is a combination of different effects which allow us to detect such high flows  which could have been blurred  without the adapative optics and the pixel resolution.

\section{Discussion and Conclusion}
We report on the morphology of jets occurring in the NOAA AR 13435  consisting of a leading spot in a decaying phase, and a large extended filament anchored in the region. The region is also close to a CH. The recurrence of the jets is explained by the continuous  emergence of moving magnetic features and cancelling flux. The deflection of the jets is due to the presence of  the CH.
The length of the jets are between 20 to 80 Mm going from West to East. There are  multiple strands with increasing plane-of-sky speeds as they reach the  open field  and are deflected towards the North. The filament visible in H$\alpha$  is surrounded by a larger channel observed  in AIA 304~\AA\ with also a deflected corridor. The recurrent jets are overlaid on the filament channel due to the perspective effect, and it is difficult to distinguish the cool plasma jet (surge) accompanying the EUV jet from the threads of the filament.

The more surprising observations come from the THEMIS spectra of one of the jet series, which indicate strong bilateral flows and strong blueshift in a small area. We propose the following scenario to explain bilateral flows of $\pm$ 80 km s$^{-1}$ and the high speed velocity (140 km s$^{-1}$). By considering that the jet structure and the filament channel have nearly the same magnetic direction West to East and overlay in some sense, interchange of magnetic field lines is possible with  magnetic reconnection between the magnetic field lines of the jet and  the filament channel. In Figure \ref{fig:jet_filament_drawing} we propose a  sketch,  with a 
 reconnection jet. Blueshift and redshift along the jet could correspond to a kind of twist of the jet. However, at the location of the bilateral flows, a brightening in H$\alpha$ is observed, indicating some heating which could be due to magnetic reconnection.  It corresponds to brightenings identified in AIA 171, 211, and 304~\AA\  pointed by black arrows in Figure \ref{fig:211evolution}. The blueshifted material could be ejected plasmoid  during the reconnection. This scenario is similar to the nanojets 
observed in loop magnetic field lines by \citealt{Antolin2021} (Figure 15 in the supplement).

To conclude, we present joint observations using high spatial resolution H$\alpha$ spectra obtained by the ground-based telescope THEMIS in Canary Islands, together with HMI and AIA data from the Solar Dynamic Observatory to study the interaction between a filament channel and recurring jets. These observations suggest the presence of magnetic field lines interchange between the filament channel and the jets, leading to cool plasmoid-ejections or 
reconnection jets with high speed flows perpendicularly to the jet trajectory.


{\bf Acknowledgments}\\
The authors are grateful to the referee for the comments and suggestions which have improved the paper.
We thank the THEMIS team for their  support during the observations. We thank the open data policy of the SDO and THEMIS teams. We thank Etienne Pariat and Aaron Peat for fruitful discussions of the jet observed with THEMIS. GK thanks the DST/INSPIRE program. PD is supported by CSIR, New Delhi. BS was support
ed by SOLARNET EU program under the project ``Magnetic field structure of prominences, solar tornadoes and spicules". R.C. acknowledges the support from DST/SERB project No. EEQ/2023/000214.
NL acknowledges support from UK Research and Innovation's Science and Technology Facilities Council under grant award numbers ST/T000422/1 and ST/X000990/1. RJ acknowledges the support by the Research Council of Norway, through its Centres of Excellence scheme, project number 262622.

\bibliographystyle{aasjournal}
\bibliography{references}

\begin{thebibliography}{}
\expandafter\ifx\csname natexlab\endcsname\relax\def\natexlab#1{#1}\fi
\providecommand{\url}[1]{\href{#1}{#1}}
\providecommand{\dodoi}[1]{doi:~\href{http://doi.org/#1}{\nolinkurl{#1}}}
\providecommand{\doeprint}[1]{\href{http://ascl.net/#1}{\nolinkurl{http://ascl.net/#1}}}
\providecommand{\doarXiv}[1]{\href{https://arxiv.org/abs/#1}{\nolinkurl{https://arxiv.org/abs/#1}}}

\bibitem[{{Alexander} \& {Fletcher}(1999)}]{Alexander1999}
{Alexander}, D., \& {Fletcher}, L. 1999, \solphys, 190, 167, \dodoi{10.1023/A:1005213826793}

\bibitem[{{Antolin} {et~al.}(2021){Antolin}, {Pagano}, {Testa}, {Petralia}, \& {Reale}}]{Antolin2021}
{Antolin}, P., {Pagano}, P., {Testa}, P., {Petralia}, A., \& {Reale}, F. 2021, Nature Astronomy, 5, 54, \dodoi{10.1038/s41550-020-1199-8}

\bibitem[{{Aulanier} \& {Demoulin}(1998)}]{1998Aulanier1998}
{Aulanier}, G., \& {Demoulin}, P. 1998, \aap, 329, 1125

\bibitem[{{Aulanier} {et~al.}(2002){Aulanier}, {DeVore}, \& {Antiochos}}]{Aulanier_De2002}
{Aulanier}, G., {DeVore}, C.~R., \& {Antiochos}, S.~K. 2002, \apjl, 567, L97, \dodoi{10.1086/339436}

\bibitem[{{Aulanier} \& {Schmieder}(2002)}]{Aulanier_BS2002}
{Aulanier}, G., \& {Schmieder}, B. 2002, \aap, 386, 1106, \dodoi{10.1051/0004-6361:20020179}

\bibitem[{{Berghmans} {et~al.}(2021){Berghmans}, {Auch{\`e}re}, {Long}, {Soubri{\'e}}, {Mierla}, {Zhukov}, {Sch{\"u}hle}, {Antolin}, {Harra}, {Parenti}, {Podladchikova}, {Aznar Cuadrado}, {Buchlin}, {Dolla}, {Verbeeck}, {Gissot}, {Teriaca}, {Haberreiter}, {Katsiyannis}, {Rodriguez}, {Kraaikamp}, {Smith}, {Stegen}, {Rochus}, {Halain}, {Jacques}, {Thompson}, \& {Inhester}}]{Berghmans2021}
{Berghmans}, D., {Auch{\`e}re}, F., {Long}, D.~M., {et~al.} 2021, \aap, 656, L4, \dodoi{10.1051/0004-6361/202140380}

\bibitem[{{Cai} {et~al.}(2024){Cai}, {Ruan}, {Zheng}, {Schmieder}, {Guo}, {Chen}, {Su}, {Liu}, {Liu}, \& {Cao}}]{Cai2024}
{Cai}, Q., {Ruan}, G., {Zheng}, C., {et~al.} 2024, \aap, 682, A183, \dodoi{10.1051/0004-6361/202348053}

\bibitem[{{Canfield} {et~al.}(1996){Canfield}, {Reardon}, {Leka}, {Shibata}, {Yokoyama}, \& {Shimojo}}]{Canfield1996}
{Canfield}, R.~C., {Reardon}, K.~P., {Leka}, K.~D., {et~al.} 1996, \apj, 464, 1016, \dodoi{10.1086/177389}

\bibitem[{{Chandra} {et~al.}(2015){Chandra}, {Gupta}, {Mulay}, \& {Tripathi}}]{Chandra2015}
{Chandra}, R., {Gupta}, G.~R., {Mulay}, S., \& {Tripathi}, D. 2015, \mnras, 446, 3741, \dodoi{10.1093/mnras/stu2305}

\bibitem[{{Chandra} {et~al.}(2017){Chandra}, {Mandrini}, {Schmieder}, {Joshi}, {Cristiani}, {Cremades}, {Pariat}, {Nuevo}, {Srivastava}, \& {Uddin}}]{Chandra2017}
{Chandra}, R., {Mandrini}, C.~H., {Schmieder}, B., {et~al.} 2017, \aap, 598, A41, \dodoi{10.1051/0004-6361/201628984}

\bibitem[{{Guo} {et~al.}(2010){Guo}, {Schmieder}, {D{\'e}moulin}, {Wiegelmann}, {Aulanier}, {T{\"o}r{\"o}k}, \& {Bommier}}]{Guo2010}
{Guo}, Y., {Schmieder}, B., {D{\'e}moulin}, P., {et~al.} 2010, \apj, 714, 343, \dodoi{10.1088/0004-637X/714/1/343}

\bibitem[{{Harvey} {et~al.}(1996){Harvey}, {Hill}, {Hubbard}, {Kennedy}, {Leibacher}, {Pintar}, {Gilman}, {Noyes}, {Title}, {Toomre}, {Ulrich}, {Bhatnagar}, {Kennewell}, {Marquette}, {Patron}, {Saa}, \& {Yasukawa}}]{Harvey1996}
{Harvey}, J.~W., {Hill}, F., {Hubbard}, R.~P., {et~al.} 1996, Science, 272, 1284, \dodoi{10.1126/science.272.5266.1284}

\bibitem[{{Heinzel} \& {Anzer}(2001)}]{Heinzel2001}
{Heinzel}, P., \& {Anzer}, U. 2001, \aap, 375, 1082, \dodoi{10.1051/0004-6361:20010926}

\bibitem[{{Jiang} {et~al.}(2007){Jiang}, {Chen}, {Li}, {Shen}, \& {Yang}}]{Jiang2007}
{Jiang}, Y.~C., {Chen}, H.~D., {Li}, K.~J., {Shen}, Y.~D., \& {Yang}, L.~H. 2007, \aap, 469, 331, \dodoi{10.1051/0004-6361:20053954}

\bibitem[{{Joshi} {et~al.}(2024{\natexlab{a}}){Joshi}, {Aulanier}, {Radcliffe}, {Rouppe van der Voort}, {Pariat}, {N{\'o}brega-Siverio}, \& {Schmieder}}]{Joshi2024}
{Joshi}, R., {Aulanier}, G., {Radcliffe}, A., {et~al.} 2024{\natexlab{a}}, \aap, 687, A172, \dodoi{10.1051/0004-6361/202449553}

\bibitem[{{Joshi} {et~al.}(2023){Joshi}, {Luna}, {Schmieder}, {Moreno-Insertis}, \& {Chandra}}]{Joshi2023}
{Joshi}, R., {Luna}, M., {Schmieder}, B., {Moreno-Insertis}, F., \& {Chandra}, R. 2023, \aap, 672, A15, \dodoi{10.1051/0004-6361/202245647}

\bibitem[{{Joshi} {et~al.}(2024{\natexlab{b}}){Joshi}, {Rouppe van der Voort}, {Schmieder}, {Moreno-Insertis}, {Prasad}, {Aulanier}, \& {N{\'o}brega-Siverio}}]{Joshi2024F}
{Joshi}, R., {Rouppe van der Voort}, L., {Schmieder}, B., {et~al.} 2024{\natexlab{b}}, \aap, 691, A198, \dodoi{10.1051/0004-6361/202449715}

\bibitem[{{Joshi} {et~al.}(2020{\natexlab{a}}){Joshi}, {Schmieder}, {Aulanier}, {Bommier}, \& {Chandra}}]{Joshi_Gui2020}
{Joshi}, R., {Schmieder}, B., {Aulanier}, G., {Bommier}, V., \& {Chandra}, R. 2020{\natexlab{a}}, \aap, 642, A169, \dodoi{10.1051/0004-6361/202038562}

\bibitem[{{Joshi} {et~al.}(2017){Joshi}, {Schmieder}, {Chandra}, {Aulanier}, {Zuccarello}, \& {Uddin}}]{Joshi2017}
{Joshi}, R., {Schmieder}, B., {Chandra}, R., {et~al.} 2017, \solphys, 292, 152, \dodoi{10.1007/s11207-017-1176-2}

\bibitem[{{Joshi} {et~al.}(2020{\natexlab{b}}){Joshi}, {Wang}, {Chandra}, {Zhang}, {Liu}, \& {Li}}]{Joshi_CME2020}
{Joshi}, R., {Wang}, Y., {Chandra}, R., {et~al.} 2020{\natexlab{b}}, \apj, 901, 94, \dodoi{10.3847/1538-4357/abaf5a}

\bibitem[{{Labrosse} {et~al.}(2010){Labrosse}, {Heinzel}, {Vial}, {Kucera}, {Parenti}, {Gun{\'a}r}, {Schmieder}, \& {Kilper}}]{Labrosse2010}
{Labrosse}, N., {Heinzel}, P., {Vial}, J.~C., {et~al.} 2010, \ssr, 151, 243, \dodoi{10.1007/s11214-010-9630-6}

\bibitem[{{Labrosse} {et~al.}(2011){Labrosse}, {Schmieder}, {Heinzel}, \& {Watanabe}}]{Labrosse2011}
{Labrosse}, N., {Schmieder}, B., {Heinzel}, P., \& {Watanabe}, T. 2011, \aap, 531, A69, \dodoi{10.1051/0004-6361/201015064}

\bibitem[{{Lemen} {et~al.}(2012){Lemen}, {Title}, {Akin}, {Boerner}, {Chou}, {Drake}, {Duncan}, {Edwards}, {Friedlaender}, {Heyman}, {Hurlburt}, {Katz}, {Kushner}, {Levay}, {Lindgren}, {Mathur}, {McFeaters}, {Mitchell}, {Rehse}, {Schrijver}, {Springer}, {Stern}, {Tarbell}, {Wuelser}, {Wolfson}, {Yanari}, {Bookbinder}, {Cheimets}, {Caldwell}, {Deluca}, {Gates}, {Golub}, {Park}, {Podgorski}, {Bush}, {Scherrer}, {Gummin}, {Smith}, {Auker}, {Jerram}, {Pool}, {Soufli}, {Windt}, {Beardsley}, {Clapp}, {Lang}, \& {Waltham}}]{Lemen2012}
{Lemen}, J.~R., {Title}, A.~M., {Akin}, D.~J., {et~al.} 2012, \solphys, 275, 17, \dodoi{10.1007/s11207-011-9776-8}

\bibitem[{{Liu} \& {Kurokawa}(2004)}]{Liu2004}
{Liu}, Y., \& {Kurokawa}, H. 2004, \apj, 610, 1136, \dodoi{10.1086/421715}

\bibitem[{{Liu} {et~al.}(2022){Liu}, {Ruan}, {Schmieder}, {Masson}, {Chen}, {Su}, {Wang}, {Bai}, {Su}, \& {Cao}}]{Liu2022}
{Liu}, Y., {Ruan}, G.~P., {Schmieder}, B., {et~al.} 2022, \aap, 667, A24, \dodoi{10.1051/0004-6361/202243292}

\bibitem[{{Luna} {et~al.}(2024){Luna}, {Joshi}, {Schmieder}, {Moreno-Insertis}, {Liakh}, \& {Terradas}}]{Luna2024}
{Luna}, M., {Joshi}, R., {Schmieder}, B., {et~al.} 2024, arXiv e-prints, arXiv:2410.10223.
\newblock \doarXiv{2410.10223}

\bibitem[{{Luna} \& {Moreno-Insertis}(2021)}]{Luna2021}
{Luna}, M., \& {Moreno-Insertis}, F. 2021, \apj, 912, 75, \dodoi{10.3847/1538-4357/abec46}

\bibitem[{{Mackay} {et~al.}(2010){Mackay}, {Karpen}, {Ballester}, {Schmieder}, \& {Aulanier}}]{Mackay2010}
{Mackay}, D.~H., {Karpen}, J.~T., {Ballester}, J.~L., {Schmieder}, B., \& {Aulanier}, G. 2010, \ssr, 151, 333, \dodoi{10.1007/s11214-010-9628-0}

\bibitem[{{Mein} \& {Rayrole}(1985)}]{Mein1985}
{Mein}, P., \& {Rayrole}, J. 1985, Vistas in Astronomy, 28, 567, \dodoi{10.1016/0083-6656(85)90077-7}

\bibitem[{{Moore} {et~al.}(2010){Moore}, {Cirtain}, {Sterling}, \& {Falconer}}]{Moore2010}
{Moore}, R.~L., {Cirtain}, J.~W., {Sterling}, A.~C., \& {Falconer}, D.~A. 2010, \apj, 720, 757, \dodoi{10.1088/0004-637X/720/1/757}

\bibitem[{{Moore} {et~al.}(2022){Moore}, {Panesar}, {Sterling}, \& {Tiwari}}]{Moore2022}
{Moore}, R.~L., {Panesar}, N.~K., {Sterling}, A.~C., \& {Tiwari}, S.~K. 2022, \apj, 933, 12, \dodoi{10.3847/1538-4357/ac6181}

\bibitem[{{Mulay} {et~al.}(2016){Mulay}, {Tripathi}, {Del Zanna}, \& {Mason}}]{Mulay2016}
{Mulay}, S.~M., {Tripathi}, D., {Del Zanna}, G., \& {Mason}, H. 2016, \aap, 589, A79, \dodoi{10.1051/0004-6361/201527473}

\bibitem[{{Nistic{\`o}} {et~al.}(2009){Nistic{\`o}}, {Bothmer}, {Patsourakos}, \& {Zimbardo}}]{Nistico2009}
{Nistic{\`o}}, G., {Bothmer}, V., {Patsourakos}, S., \& {Zimbardo}, G. 2009, \solphys, 259, 87, \dodoi{10.1007/s11207-009-9424-8}

\bibitem[{{N{\'o}brega-Siverio} {et~al.}(2021){N{\'o}brega-Siverio}, {Guglielmino}, \& {Sainz Dalda}}]{Nobrega-Siverio2021}
{N{\'o}brega-Siverio}, D., {Guglielmino}, S.~L., \& {Sainz Dalda}, A. 2021, \aap, 655, A28, \dodoi{10.1051/0004-6361/202141472}

\bibitem[{{N{\'o}brega-Siverio} {et~al.}(2017){N{\'o}brega-Siverio}, {Mart{\'\i}nez-Sykora}, {Moreno-Insertis}, \& {Rouppe van der Voort}}]{Nobrega-Siverio2017}
{N{\'o}brega-Siverio}, D., {Mart{\'\i}nez-Sykora}, J., {Moreno-Insertis}, F., \& {Rouppe van der Voort}, L. 2017, \apj, 850, 153, \dodoi{10.3847/1538-4357/aa956c}

\bibitem[{{Panesar} {et~al.}(2016{\natexlab{a}}){Panesar}, {Sterling}, \& {Moore}}]{Panesar2016b}
{Panesar}, N.~K., {Sterling}, A.~C., \& {Moore}, R.~L. 2016{\natexlab{a}}, \apjl, 822, L23, \dodoi{10.3847/2041-8205/822/2/L23}

\bibitem[{{Panesar} {et~al.}(2016{\natexlab{b}}){Panesar}, {Sterling}, {Moore}, \& {Chakrapani}}]{Panesar2016a}
{Panesar}, N.~K., {Sterling}, A.~C., {Moore}, R.~L., \& {Chakrapani}, P. 2016{\natexlab{b}}, \apjl, 832, L7, \dodoi{10.3847/2041-8205/832/1/L7}

\bibitem[{{Panesar} {et~al.}(2021){Panesar}, {Tiwari}, {Berghmans}, {Cheung}, {M{\"u}ller}, {Auchere}, \& {Zhukov}}]{Panesar2021}
{Panesar}, N.~K., {Tiwari}, S.~K., {Berghmans}, D., {et~al.} 2021, \apjl, 921, L20, \dodoi{10.3847/2041-8213/ac3007}

\bibitem[{{Paraschiv} {et~al.}(2015){Paraschiv}, {Bemporad}, \& {Sterling}}]{Paraschiv2015}
{Paraschiv}, A.~R., {Bemporad}, A., \& {Sterling}, A.~C. 2015, \aap, 579, A96, \dodoi{10.1051/0004-6361/201525671}

\bibitem[{{Pesnell} {et~al.}(2012){Pesnell}, {Thompson}, \& {Chamberlin}}]{Pesnell2012}
{Pesnell}, W.~D., {Thompson}, B.~J., \& {Chamberlin}, P.~C. 2012, \solphys, 275, 3, \dodoi{10.1007/s11207-011-9841-3}

\bibitem[{{Poisson} {et~al.}(2020){Poisson}, {Bustos}, {L{\'o}pez Fuentes}, {Mandrini}, \& {Cristiani}}]{Poisson2020}
{Poisson}, M., {Bustos}, C., {L{\'o}pez Fuentes}, M., {Mandrini}, C.~H., \& {Cristiani}, G.~D. 2020, Advances in Space Research, 65, 1629, \dodoi{10.1016/j.asr.2019.09.026}

\bibitem[{{Raouafi} {et~al.}(2016){Raouafi}, {Patsourakos}, {Pariat}, {Young}, {Sterling}, {Savcheva}, {Shimojo}, {Moreno-Insertis}, {DeVore}, {Archontis}, {T{\"o}r{\"o}k}, {Mason}, {Curdt}, {Meyer}, {Dalmasse}, \& {Matsui}}]{Raouafi2016}
{Raouafi}, N.~E., {Patsourakos}, S., {Pariat}, E., {et~al.} 2016, \ssr, 201, 1, \dodoi{10.1007/s11214-016-0260-5}

\bibitem[{{Roy}(1973)}]{Roy1973}
{Roy}, J.~R. 1973, \solphys, 28, 95, \dodoi{10.1007/BF00152915}

\bibitem[{{Ruan} {et~al.}(2019){Ruan}, {Schmieder}, {Masson}, {Mein}, {Mein}, {Aulanier}, \& {Chen}}]{Ruan2019}
{Ruan}, G., {Schmieder}, B., {Masson}, S., {et~al.} 2019, \apj, 883, 52, \dodoi{10.3847/1538-4357/ab3657}

\bibitem[{{Schmieder} {et~al.}(2022){Schmieder}, {Joshi}, \& {Chandra}}]{Schmieder2022}
{Schmieder}, B., {Joshi}, R., \& {Chandra}, R. 2022, Advances in Space Research, 70, 1580, \dodoi{10.1016/j.asr.2021.12.013}

\bibitem[{{Schmieder} {et~al.}(2004){Schmieder}, {Lin}, {Heinzel}, \& {Schwartz}}]{Schmieder_Lin2004}
{Schmieder}, B., {Lin}, Y., {Heinzel}, P., \& {Schwartz}, P. 2004, \solphys, 221, 297, \dodoi{10.1023/B:SOLA.0000035059.50427.68}

\bibitem[{{Schmieder} {et~al.}(1996{\natexlab{a}}){Schmieder}, {Mein}, {Shibata}, {van Driel-Gesztelyi}, \& {Kurokawa}}]{Schmieder_Shi1996}
{Schmieder}, B., {Mein}, N., {Shibata}, K., {van Driel-Gesztelyi}, L., \& {Kurokawa}, H. 1996{\natexlab{a}}, Advances in Space Research, 17, 193, \dodoi{10.1016/0273-1177(95)00566-W}

\bibitem[{{Schmieder} {et~al.}(1984){Schmieder}, {Mein}, {Martres}, \& {Tandberg-Hanssen}}]{Schmieder1984}
{Schmieder}, B., {Mein}, P., {Martres}, M.~J., \& {Tandberg-Hanssen}, E. 1984, \solphys, 94, 133, \dodoi{10.1007/BF00154814}

\bibitem[{{Schmieder} {et~al.}(1988){Schmieder}, {Mein}, {Simnett}, \& {Tandberg-Hanssen}}]{Schmieder1988}
{Schmieder}, B., {Mein}, P., {Simnett}, G.~M., \& {Tandberg-Hanssen}, E. 1988, \aap, 201, 327

\bibitem[{{Schmieder} {et~al.}(1983){Schmieder}, {Mein}, {Vial}, \& {Tandberg-Hanssen}}]{Schmieder1983}
{Schmieder}, B., {Mein}, P., {Vial}, J.~C., \& {Tandberg-Hanssen}, E. 1983, \aap, 127, 337

\bibitem[{{Schmieder} {et~al.}(1996{\natexlab{b}}){Schmieder}, {Rovira}, {Simnett}, {Fontenla}, \& {Tandberg-Hanssen}}]{Schmieder_Sim1996}
{Schmieder}, B., {Rovira}, M., {Simnett}, G.~M., {Fontenla}, J.~M., \& {Tandberg-Hanssen}, E. 1996{\natexlab{b}}, \aap, 308, 957

\bibitem[{{Schmieder} {et~al.}(2013){Schmieder}, {Guo}, {Moreno-Insertis}, {Aulanier}, {Yelles Chaouche}, {Nishizuka}, {Harra}, {Thalmann}, {Vargas Dominguez}, \& {Liu}}]{Schmieder2013}
{Schmieder}, B., {Guo}, Y., {Moreno-Insertis}, F., {et~al.} 2013, \aap, 559, A1, \dodoi{10.1051/0004-6361/201322181}

\bibitem[{{Schou} {et~al.}(2012){Schou}, {Scherrer}, {Bush}, {Wachter}, {Couvidat}, {Rabello-Soares}, {Bogart}, {Hoeksema}, {Liu}, {Duvall}, {Akin}, {Allard}, {Miles}, {Rairden}, {Shine}, {Tarbell}, {Title}, {Wolfson}, {Elmore}, {Norton}, \& {Tomczyk}}]{Schou2012}
{Schou}, J., {Scherrer}, P.~H., {Bush}, R.~I., {et~al.} 2012, \solphys, 275, 229, \dodoi{10.1007/s11207-011-9842-2}

\bibitem[{{Shen}(2021)}]{Shen2021}
{Shen}, Y. 2021, Proceedings of the Royal Society of London Series A, 477, 217, \dodoi{10.1098/rspa.2020.0217}

\bibitem[{{Shibata} {et~al.}(1994){Shibata}, {Nitta}, {Matsumoto}, {Tajima}, {Yokoyama}, {Hirayama}, \& {Hudson}}]{Shibata1994}
{Shibata}, K., {Nitta}, N., {Matsumoto}, R., {et~al.} 1994, in X-ray solar physics from Yohkoh, ed. Y.~{Uchida}, T.~{Watanabe}, K.~{Shibata}, \& H.~S. {Hudson}, 29

\bibitem[{{Shibata} {et~al.}(1992){Shibata}, {Ishido}, {Acton}, {Strong}, {Hirayama}, {Uchida}, {McAllister}, {Matsumoto}, {Tsuneta}, {Shimizu}, {Hara}, {Sakurai}, {Ichimoto}, {Nishino}, \& {Ogawara}}]{Shibata1992}
{Shibata}, K., {Ishido}, Y., {Acton}, L.~W., {et~al.} 1992, \pasj, 44, L173

\bibitem[{{Shimojo} {et~al.}(1996){Shimojo}, {Hashimoto}, {Shibata}, {Hirayama}, {Hudson}, \& {Acton}}]{Shimojo1996}
{Shimojo}, M., {Hashimoto}, S., {Shibata}, K., {et~al.} 1996, \pasj, 48, 123, \dodoi{10.1093/pasj/48.1.123}

\bibitem[{{Sterling} {et~al.}(2015){Sterling}, {Moore}, {Falconer}, \& {Adams}}]{Sterling2015}
{Sterling}, A.~C., {Moore}, R.~L., {Falconer}, D.~A., \& {Adams}, M. 2015, \nat, 523, 437, \dodoi{10.1038/nature14556}

\bibitem[{{Sterling} {et~al.}(2016){Sterling}, {Moore}, {Falconer}, {Panesar}, {Akiyama}, {Yashiro}, \& {Gopalswamy}}]{Sterling2016}
{Sterling}, A.~C., {Moore}, R.~L., {Falconer}, D.~A., {et~al.} 2016, \apj, 821, 100, \dodoi{10.3847/0004-637X/821/2/100}

\bibitem[{{Sterling} {et~al.}(2022){Sterling}, {Moore}, \& {Panesar}}]{Sterling2022}
{Sterling}, A.~C., {Moore}, R.~L., \& {Panesar}, N.~K. 2022, \apj, 927, 127, \dodoi{10.3847/1538-4357/ac473f}

\bibitem[{{Tallon} {et~al.}(2022){Tallon}, {Thi{\'e}baut}, {Tallon-Bosc}, {Gelly}, \& {Denis}}]{Tallon2022}
{Tallon}, M., {Thi{\'e}baut}, {\'E}., {Tallon-Bosc}, I., {Gelly}, B., \& {Denis}, L. 2022, in Society of Photo-Optical Instrumentation Engineers (SPIE) Conference Series, Vol. 12185, Adaptive Optics Systems VIII, ed. L.~{Schreiber}, D.~{Schmidt}, \& E.~{Vernet}, 121852J, \dodoi{10.1117/12.2630379}

\bibitem[{{T{\"o}r{\"o}k} {et~al.}(2009){T{\"o}r{\"o}k}, {Aulanier}, {Schmieder}, {Reeves}, \& {Golub}}]{Torok2009}
{T{\"o}r{\"o}k}, T., {Aulanier}, G., {Schmieder}, B., {Reeves}, K.~K., \& {Golub}, L. 2009, \apj, 704, 485, \dodoi{10.1088/0004-637X/704/1/485}

\bibitem[{{Uddin} {et~al.}(2012){Uddin}, {Schmieder}, {Chandra}, {Srivastava}, {Kumar}, \& {Bisht}}]{Uddin2012}
{Uddin}, W., {Schmieder}, B., {Chandra}, R., {et~al.} 2012, \apj, 752, 70, \dodoi{10.1088/0004-637X/752/1/70}

\bibitem[{{Wyper} {et~al.}(2018){Wyper}, {DeVore}, \& {Antiochos}}]{Wyper2018}
{Wyper}, P.~F., {DeVore}, C.~R., \& {Antiochos}, S.~K. 2018, \apj, 852, 98, \dodoi{10.3847/1538-4357/aa9ffc}

\bibitem[{{Wyper} {et~al.}(2019){Wyper}, {DeVore}, \& {Antiochos}}]{Wyper2019}
---. 2019, \mnras, 490, 3679, \dodoi{10.1093/mnras/stz2674}

\bibitem[{{Zhang} {et~al.}(2023){Zhang}, {Musset}, {Glesener}, {Panesar}, \& {Fleishman}}]{Zhang2023}
{Zhang}, Y., {Musset}, S., {Glesener}, L., {Panesar}, N.~K., \& {Fleishman}, G.~D. 2023, \apj, 943, 180, \dodoi{10.3847/1538-4357/aca654}

\bibitem[{{Zhukov} {et~al.}(2021){Zhukov}, {Mierla}, {Auch{\`e}re}, {Gissot}, {Rodriguez}, {Soubri{\'e}}, {Thompson}, {Inhester}, {Nicula}, {Antolin}, {Parenti}, {Buchlin}, {Barczynski}, {Verbeeck}, {Kraaikamp}, {Smith}, {Stegen}, {Dolla}, {Harra}, {Long}, {Sch{\"u}hle}, {Podladchikova}, {Aznar Cuadrado}, {Teriaca}, {Haberreiter}, {Katsiyannis}, {Rochus}, {Halain}, {Jacques}, \& {Berghmans}}]{Zhukov2021}
{Zhukov}, A.~N., {Mierla}, M., {Auch{\`e}re}, F., {et~al.} 2021, \aap, 656, A35, \dodoi{10.1051/0004-6361/202141010}

\end{thebibliography}


\end{document}